\documentclass[prodmode,acmtist]{acmsmall}

\usepackage[ruled]{algorithm2e}

\SetAlFnt{\small}
\SetAlCapFnt{\small}
\SetAlCapNameFnt{\small}
\SetAlCapHSkip{0pt}
\IncMargin{-\parindent}

\usepackage{graphicx}
\usepackage{url}
\usepackage{subfigure}
\usepackage{booktabs}
\usepackage{multirow}
\usepackage{tabularx}

\acmVolume{5}
\acmNumber{1}
\acmArticle{1}
\acmYear{2014}
\acmMonth{1}

\begin{document}

\markboth{P. De Meo et al.}{Analyzing User Behavior across Social Sharing Environments}

\title{Analyzing User Behavior across Social Sharing Environments}

\author{
Pasquale De Meo
\affil{University of Messina}
Emilio Ferrara
\affil{Indiana University Bloomington, USA}
Fabian Abel
\affil{TU Delft}
Lora Aroyo
\affil{VU University Amsterdam}
Geert-Jan Houben
\affil{TU Delft}
}

\begin{abstract}
In this work we present an in-depth analysis of the user behaviors on different Social Sharing systems. We consider three popular platforms, \emph{Flickr}, \emph{Delicious} and \emph{StumbleUpon}, and, by combining techniques from social network analysis with techniques from
semantic analysis, we characterize the tagging behavior as well as the tendency to create friendship relationships of the users
of these platforms. The aim of our investigation is to see if (and how) the features and goals of a given Social Sharing system reflect on the behavior of its users and, moreover, if there exists a correlation between the social and tagging behavior of the users. We report our findings in terms of the characteristics of user profiles according to three different dimensions: {\em (i)} intensity of user activities, {\em (ii)} tag-based characteristics of user profiles, and {\em (iii)} semantic characteristics of user profiles.

\end{abstract}




\acmformat{De Meo, P., Ferrara, E., Abel, F., Aroyo, L., Houben, G.-J. Analyzing User
Behavior across Social Sharing Environments.}

\begin{bottomstuff}
Author's addresses: P. De Meo - Department of Ancient and Modern Civilizations, University of Messina, Italy, {\tt
pdemeo@unime.it}, E. Ferrara - Center for Complex Networks and Systems Research
School of Informatics and Computing, Indiana University Bloomington, US,
{\tt ferrarae@indiana.edu}, F. Abel - Department of Software and Computer Technology, Web Information Systems Group, Technical
University of Delft, The Netherlands, {\tt f.abel@tudelft.nl}, L. Aroyo - Department of
Computer Science, VU University Amsterdam, The Netherlands, {\tt l.m.aroyo@vu.nl}, G.-J.
Houben - Department of Software and Computer Technology, Web Information Systems Group, Technical University of Delft, The
Netherlands, {\tt g.j.p.m.houben@tudelft.nl}.
\end{bottomstuff}

\maketitle

\section{Introduction}
\label{sec:intro}

The unprecedented and quickly expanding success of the Social Web and related
phenomena such as  \emph{social networking} and \emph{collaborative tagging}, raised the importance
of studying the behavior and the habits of Social Web users.

Several authors were committed to studying human behaviors within each single platform, considered as a {\em single} and {\em autonomous entity} \cite{ferrara2011crawling}.

On top of that, investigating user behavior is even more interesting across systems, if we consider
the fact that, nowadays, users have a plethora of possible choices and often they subscribe
to different platforms that provide different but related services \cite{francesca/UserIdentification,DBLP:conf/ideas/MeoNQRU09,tereza/identifyingUserViaTags}.
Indeed, users can behave on each of those platforms in similar (or different) manners, depending on the context and the platform.

In this paper we report our findings on the behavior of users belonging to different social platforms.
We focused on {\em Social Sharing systems}, i.e., on Social Web
systems allowing users to post, tag and share resources as well as to become friends each other.
In our research, we considered three popular social platforms, namely \emph{Flickr},
\emph{Delicious} and \emph{StumbleUpon}.

We decided to consider these systems for two main reasons: first of all, they are quite popular and this allowed us to gather a sufficiently
large data sample to perform a significant analysis. Secondly, in these systems, user contributed tags are a powerful {\em knowledge management tool}: tags, in fact, can be used to make the retrieval of resources easier or to raise the visibility of user generated contents. Tags, in other words, have also a {\em semantic value} because from their analysis it is possible to infer general concepts exploited by users \cite{cantador2011categorising,de2009exploitation}. Tagging is also present in other platforms but with different purposes and
goals: for instance, in Facebook users are allowed to apply tags but their tagging activity has
not a semantic value: a tag in Facebook is used to link objects (e.g., a person to a photo). Therefore, if a
user tags people, pages and places in her posts, she has the goal, for instance, to
inform her friends about the people she is with.

In our investigation we focused on two, possibly related, aspects, i.e., \emph{(i)}
understanding to what extent the architecture and the features of a Social Sharing system
influence the behavior of its users, and, \emph{(ii)} analyzing if there exists a relation
between the users' social behavior (i.e., the tendency of users to socialize among each other) and their
tagging behavior (i.e., the aptitude of users to label resources).

Regarding point \emph{(i)}, we investigate \emph{(a)} if the number of friends of a user significantly varies across different systems; \emph{(b)} if users exploit a similar amount of tags and, from a semantic perspective, if users in some systems prefer to focus on a narrow set of topics whereas in other systems their interests are described by a broader group of topics; \emph{(c)} if users exploit each platform to focus on a specific topic or, otherwise,
if they discuss similar topics on different platforms.

As for point \emph{(ii)}, we investigate \emph{(d)} if there exists a correlation between the activity of tagging resources and the tendency to making new friends across the given Social Sharing platforms, and, most importantly, if the behavior of a given user is somehow similar to the behavior of her friends. To this purpose we studied \emph{(e)} the semantic variety of topics exploited by a user against those addressed by her friends, and, \emph{(f)} the breadth of the topics covered by a user against those exploited by her contacts.

To do so, in this work we adopt techniques from \emph{social network analysis} and {\em semantic-based analysis}. In particular, we studied at the statistical level the frequency of tagging activities and the frequency of friending activities. We used external
thesauri (like {\em WordNet} \cite{Fellbaum1998}) and ontologies (like {\em DBPedia} \cite{dbpedia/iswc07}) to map user contributed tags onto lexicographic or
ontological concepts and infer the topics exploited by users.

The importance and relevance of investigating these aspects is motivated by a multitude of
scientific, economic and societal considerations.

For example, understanding if the tagging and social behavior of a given user on a specific platform depends on the
architecture and design feature of the platform itself could help platform designers to decide whether to modify the ways users interact with the platform in order to maximize, for instance, their loyalty.

From a sociological perspective, if we would prove that users show a particular social behavior on a given platform, this could be helpful to design novel social theories explaining how user relationships emerge and evolve over time depending on the design features of the platform a user is currently accessing.

The main findings of our work can be summarized as follows:

\begin{enumerate}

\item The tendency of a user to socialize depends on the features of the system in which she operates in.

\item In Delicious and Flickr, the aptitude of a user to apply tags is independent of the features of these systems. By contrast, in StumbleUpon, the features of the system have a major impact on the tagging behavior of a user.

\item The vast majority of users tend to apply most of their tags in a uniform fashion
    independently of the Social Sharing platform in which they operate in.

\item In Flickr there is a strong correlation between the intensity of tagging
    and the intensity of friending activities. By contrast, in Delicious the intensity of the tagging activity is
    much more marked than the intensity of friending activities.

\item The variety of lexicographic concepts that people use within different Social Sharing systems does not sensibly vary.
    However, the variety of ontological concepts that people exploit strongly differs from a system to another one.

\end{enumerate}

The rest of the paper is organized as follows.
In Section \ref{sec:background} we present some background material as well as a detailed description of the main features of Flickr, Delicious and StumbleUpon. Section \ref{sec:researchquestions} introduces our research methodology and, in particular, we sketched some of the research questions that this work aims to address. The datasets exploited to perform our analysis are illustrated in Section \ref{sec:datasets}.
To answer our research questions, we analyze user profiles at three different levels:  \emph{intensity level} (discussed in Section \ref{sub:intensity}), \emph{tag level} (analyzed in Section \ref{sub:userprofiletagvariety}), and \emph{semantic level} (covered in Section \ref{sec:userprofile-semantics}). In Section \ref{sec:applications} we illustrate some practical applications of our findings.
We discuss related literature in Section \ref{sec:related}. Finally, Section \ref{sec:conclusions} concludes the paper, summarizing our most important findings and depicting some future directions of research.

\section{Background}
\label{sec:background}

In this section, we first introduce some formal definitions that will be extensively used throughout the paper (see Section \ref{sub:general-definitions}).
After that, in Section \ref{sub:social-systems-description} we will describe the main features of Flickr, Delicious and StumbleUpon.

\subsection{Definitions}
\label{sub:general-definitions}

Our research focuses on {\em Social Sharing platforms} in which a user is allowed to perform {\em
various activities} like becoming friends with other users, posting comments/reviews, rating
objects, joining groups of interest and so on. In the following, we refer to a Social Sharing platform
as $\mathcal{S}$ and assume that users are free to join as many Social Sharing platforms as they
want.

We focus on two specific activities: the first is called {\em tagging} and denotes
the fact that users may apply keywords ({\em tags}) to label resources. Tagging is a
crucial activity to collaboratively classify contents in the context of Web 2.0: for instance, labeling multimedia objects
like photos/videos makes it easier to retrieve them later. In addition, users can promote the
contents they produce by tagging and exposing them to other users. The usage of
tags is also relevant from a scientific standpoint: in fact, external thesauri (like WordNet) or
ontologies (like DBPedia) can be applied to determine the meaning of tags.

The results of the analysis of tagging can be partially generalized to other kind of social
activities like commenting/reviewing or rating objects. In fact, the analysis of user
comments/ratings is useful to determine to what extent a resource is able to fit user
needs. By contrast, the usage of tags implicitly denotes user interests: if a user labels an item
we may guess that this item captures her interest but we {\em can not} directly quantify the
level of interest, as we could do if the user would rate that item.

The second activity is called {\em friending} and indicates the ability of two users of becoming
friends. As for friending activity, we consider only friendship relationships that are {\em
explicitly declared} by two users. Of course, there is a possibly wide range of {\em implicit
social relationships} binding two users: think of interactions among unknown users in Question and
Answering communities.

Friending is perhaps the most popular example of social behavior in Social Sharing system. However, there are several other examples of social behaviors worth of being considered and, among them, we cite the fact that users can form groups of interest and join groups set up by other users. The analysis of dynamics of group membership has recently attracted the interest of many authors: for instance, the study presented in \cite{Vasuki*11} shows that, in a Social Web system, the user social network and the affiliation network (consisting of users and their affiliations to groups) co-evolve and, therefore, information on user friendships and past affiliations to groups is useful to predict what groups a target user is likely to join in the future. In \cite{yuan2011factorization} the authors used information on group membership in combination with Collaborative Filtering techniques to predict the rating a user would assign to an unseen item. Experiments on a real-life dataset showed that group membership is more effective than friendship in producing accurate recommendations.

The structure emerging from a collaborative tagging activity is usually known in the literature as
{\em folksonomy} \cite{iswc05/OntologiesArUs/Mika}.

In the same spirit, the structure emerging from the overall friending actions in a
Social Sharing system generates a {\em user social network}. The nodes of this social network are the users whereas each edge specifies a friendship relationship.

%
The folksonomy identifies the {\em Content Dimension} of a Social Sharing system: its analysis is useful to detect the contents and concepts generated within the system. The user social network, by contrast, identifies the {\em Social Dimension} of a Social Sharing system: its analysis is useful to shed light on the fabric of social relationships emerging in the system itself.

The Content and Social dimensions are not to be considered as {\em independent entities}: there is a strong relationship between the social behavior of the users and the tags they
generate. Such a relationship was studied for the first time in \cite{iswc05/OntologiesArUs/Mika}.
In that paper, the author proposes a model capturing the relationships linking users, resources and tags applied by
users to label resources. Other approaches used \emph{social network analysis} tools to interpret the
semantics of tags \cite{YeGiSh07}. Approaches based on social network analysis are fascinating because they put into evidence the associative and
participatory nature of tagging process because they depict users as autonomous entities who
organize knowledge according to their own rules and negotiate the meaning of tags.

%

We are now able to provide a more formal definition of the concepts outlined above. In detailed, a
folksonomy is defined as follows \cite{hotho2006ESWCFolkRank}:

\begin{definition}[Folksonomy]\label{def:folksonomy}
A \emph{folksonomy} $\mathbb{FO}$ is a qua\-dru\-ple $\mathbb{FO} = \langle U, T, R,$ $Y \rangle$,
where $U$, $T$, $R$ are finite sets of instances of {\em users}, {\em tags}, and {\em resources}
and $Y$ defines a relation (called \emph{tag assignment}) among these sets, that is, $Y \subseteq U
\times T \times R$. The relation $Y$ can be enriched with a {\em timestamp} indicating \emph{when}
the tag assignment was performed.
\end{definition}

Along similar lines, the concept of user social network is defined as follows:

\begin{definition}[User Social Network]\label{def:friendonomy}
A \emph{User Social Network} $\mathbb{SN}$ is a pair $\mathbb{SN} = \langle U, F \rangle$, where
$U$ is a finite set of instances of {\em users} and $F$  defines a relation (called
\emph{friendship assignment}) $F \subseteq U \times U $. The relation $F$ can be enriched with a
{\em timestamp} indicating \emph{when} the friendship was created.
\end{definition}

Due to Definitions \ref{def:folksonomy} and \ref{def:friendonomy}, a Social Sharing system $\mathcal{S}$ is a
pair $\mathcal{S} = \langle \mathbb{FO}, \mathbb{SN} \rangle$. In order to handle tag
and friendship assignments in a uniform fashion, we introduce the concept of {\em social object} as the atomic component of a Social Sharing system. Such a concept will be useful to define the content-based and the social-based
profile of a user as a collection of social objects. More formally:

\begin{definition}[Social Object]\label{def:social-object}
Let $\mathcal{S} = \langle \mathbb{FO}, \mathbb{SN} \rangle$ be a Social Sharing system being
$\mathbb{FO} = \langle U, T, R,$ $Y \rangle$ and $\mathbb{SN} = \langle U, F \rangle$ the
folksonomy and the user social network associated with $\mathcal{S}$. A {\em social object} $so$ is either an element
$so \in U$ or $so \in T$. A set of social objects is said to be {\em homogeneous} if all the social objects in the set
represent a tag or identify a user.
\end{definition}

The set of social objects existing in a Social Sharing system will be called {\em social object space} and
will be denoted as $\mathbb{SO}$.

Once we have defined the concept of social object, we can introduce the concept of {\em
meta-profile} of a user $u$ in a Social Sharing system $\mathcal{S}$. The meta-profile
of $u$ is an abstract concept which can be specialized to represent the content-based (or tag-based) and the
social-based profile of $u$. The formal definition of meta-profile is as follows:

\begin{definition}[Meta-profile]\label{def:user-profile}
Let: {\em (i)} $\mathcal{S} = \langle \mathbb{FO}, \mathbb{SN} \rangle$ be a Social Sharing system, {\em (ii)} $u$ be
a user in $\mathcal{S}$ and {\em (iii)} $\mathbb{C}_u = \{so_1, \ldots so_m\}$ be a set of homogeneous social
objects. Let $w : \mathbb{SO} \rightarrow \mathbf{R}$ be a function (called {\em weighting function})
mapping social objects onto real numbers. The {\em meta-profile} $MP(u)$ of $u$ is a set of pairs
\begin{equation}
MP(u)=\{\langle so_i, w(so_i)\rangle | so_i \in \mathbb{C}_u \} \nonumber
\end{equation}

If all the social objects in $\mathbb{C}_u $ are tags, then the meta-profile of $u$ is called {\em content} (or {\em tag-based}) {\em profile} of $u$ and it is denoted as $P_T(u)$. If all the social objects in $\mathbb{C}_u $ are user identifiers, then the meta-profile of $u$ is called {\em social profile} of $u$ and it is denoted as $P_S(u)$.

\end{definition}

Definition \ref{def:user-profile} is general enough to contain some popular definition of user profile in Social Web systems.
In detail, if we assume that $\mathbb{C}_u$ coincides with the set of tags applied by $u$ and, for
each tag $t_i \in \mathbb{C}_u$, we set $w(t_i)$ equal to the frequency of assigning $t_i$,
we get the definition of content-based profile already introduced in \cite{TsMaSc08}. By contrast, if $\mathbb{C}_u$ is the set
of tags exploited by $u$ to look for resources, then we re-obtain the definition of tag-based
user profile introduced in \cite{DeQuUr10}. A special case occurs if $\mathbb{C}_u$ is
the set of tags applied by $u$ and $w(t_i)$ is set equal to 1 {\em for all} $t_i \in \mathbb{C}_u$.
In such a case we obtain the well-known concept of {\em personomy} \cite{hotho2006ESWCFolkRank}.

In an analogous fashion, if we assume that $\mathbb{C}_u$ is a set of user identifiers and, for each user $v \in
\mathbb{C}_u$, we set $w(v)=1$ if $u$ and $v$ are friends and 0 otherwise, then the social profile of $u$ coincides with the network of her friends. Of course, other more complex choices are also possible: given a user $v \in \mathbb{C}_u$ we can set $w(v)$ equal to the strength of the tie joining $u$ and $v$. Such a weight can be computed, for instance, by applying the methodology outlined in \cite{GiKa09}.

The content-based and social-based profiles of a user can be studied according to different
perspectives.


As for content-based profiles, we can study them at the {\em intensity level} (e.g., the frequency of usage of tags), at the {\em variety level} (e.g., the number of distinct tags composing the profile) and, finally at the {\em topic level} (i.e., we can use external ontologies to map tags onto concepts and compute if a user is always concerned with few topics or not). By contrast, since no semantics can be assigned to friendship relationships, the social-based profile can be analyzed
only at the {\em intensity level}.

\subsection{Social System Description}
\label{sub:social-systems-description}

In this section we describe the main features of three social systems which have been exploited to carry out our analysis, namely { \em Flickr}, {\em Delicious} and {\em StumbleUpon}.

Flickr is a Web site providing its users with photo and video hosting facilities. In August 2011, the number of Flickr users amounted to more than 51 million and more than 6 billion images were available in the platform.

Delicious is a {\em social bookmarking systems}, i.e., it allows its members to label and share URLs. In the late 2008 the number of users registered to Delicious was about 5.3 million and the estimated number of available URLs was about 180 million.

StumbleUpon is a Web search engine designed with the goal of finding and recommending Web contents (like photos or Web pages) to its users. StumbleUpon relies on collaborative filtering algorithms to build communities of like-minded users. In this fashion, a user should stumble upon pages which have been explicitly recommended by her friends or peers. In April 2012, StumbleUpon announced that it had more than 25 million registered users. At the moment of collecting data, StumbleUpon did not provided social features to its users but only tagging ones.

The systems described above can be classified according to different criteria, related to how tagging activity is conceived and to the social behaviors they support.

Here we provide some classification criteria which are, in our opinion, useful to better explain user behaviors.

\begin{enumerate}

\item {\em User Social Network}. A first classification criterium depends on the fact that the user social network (see Definition \ref{def:friendonomy}) is a {\em directed} or an {\em undirected} graph. In Flickr, links between members do not require mutual approval and, therefore, the user social network in Flickr is a {\em directed graph}. In Delicious, many of the features generally available in social networking platforms like Facebook are missing (e.g.,. the capability of directly chatting with friends). However, a user $u$ is allowed to search for a specific user $v$ by means of her username. Once $v$ have been found, she can be added to the personal network of $u$; the bookmarks selected by $v$ will be streamed onto the Delicious page of $u$ and vice versa. In this way, the user social network is an \emph{undirected graph}.

\item {\em Supporting social activities beyond friendship}. One of the most relevant features of Flickr is given by the possibility of creating {\em thematic groups} reflecting interests shared by a plenty of users. In Delicious the concept of Flickr group is replaced by {\em bundles}: if the personal network of friends of a Delicious user is quite large, a user is allowed to break it into small groups (e.g., a group for friends and a group associated with colleagues)

\item {\em Free Tagging}. In both Delicious and Flickr users are allowed to decide what tags they can use. In StumbleUpon, users can both exploit pre-defined categories (like {\tt music}) as well as to provide custom tags.

\item {\em Broad vs. narrow Folksonomy}. Folksonomies can be classified into {\em broad} and {\em narrow} \cite{broadAnNarrowFolksonomy/vanderwal}: in broad folksonomies, a resource can be tagged by multiple users and every user can label a resource by applying her own tag. This implies that many tags can be associated with a single resources. Delicious is an example of broad folksonomy. By contrast, in narrow folksonomies, a resource is labeled by a single user (or, sometimes, by few users) who provide some tags that are later exploited to retrieve the resource itself. Therefore, in narrow folksonomies, tags are directly associated with a resource. Narrow folksonomies provide relevant benefits in retrieving objects which could be hard to find or that can be retrieved only if a textual description is associated with the resource. A nice example of narrow folksonomy is given by Flickr.

\end{enumerate}

In Table \ref{table:social-system-comparison} we report the main features of Flickr, Delicious and StumbleUpon according to the criteria introduced above.

\begin{table}[t]
\begin{center}
\tbl{The main features of Flickr, Delicious, and StumbleUpon platforms\label{table:social-system-comparison}
}{
\small
\begin{tabular}{|l|c|c|c|}
\hline
\textbf{Feature/System} & {\em Flickr} & {\em Delicious} & {\em StumbleUpon}  \\

\hline
\hline
Friending & Asymmetric& Symmetric& Not Available   \\
\hline
Social Features & Groups& Bundles & Not Available   \\
\hline
Free Tagging/Category-Based Tagging & Free & Free & Category-Based + Free \\
\hline
Broad/Narrow Folksonomy & Narrow & Broad & Not Available \\
\hline
\end{tabular}
}
\end{center}
\end{table}

\section{Research Methodology}
\label{sec:researchquestions}

As pointed out in the Introduction, in our research we are interested in studying two major phenomena, i.e.: {\em (i)} to investigate
to what extent the features of a Social Sharing platform influence user behaviors, and {\em (ii)} to study whether a form of
correlation exists between the social and tagging behavior of a user. This leads to
define some research questions and, for each question, to formulate a research hypothesis.

In the following we explain our research questions and related hypotheses.

\subsection{Influence of the features of a Social Sharing system on user behaviors}

A first research theme we want to explore is about the influence that the features of a Social Sharing system has on its users and, in detail, we want to check whether meaningful differences in user behaviors emerge when users shift among systems. Such an analysis leads us to formulate the following research questions and hypotheses:

\begin{itemize}

\item $Q_1$. {\em Does the intensity of user activities carried out by a user in a Social
    Sharing system $\mathcal{S}$ depend on the features of the platform?} We are interested to study to what extent the design features of a Social Sharing system stimulate users to socialize with other users and label contents.

\item $H_1$. {\em The level of user involvement depends on the features of the Social Sharing system in which
    she operates in}. We hypothesize that the features of a Social Sharing system deeply impact of the tendency of a user to create relationships and label resources.

\item $Q_2$. {\em Does the variety of tags exploited by a user depend on the features of the Social Sharing
    system in which she is operating in?} In other words, do users tend to re-use the same
    tags in different systems or, vice versa, does the process of selecting tags depend on the Social Sharing system?

\item $H_2$. {\em Users tend, on average, to use different tags}. We believe that tags are
    reliable indicators of user needs and preferences. Since users join different platforms with
    different goals, the tags they use in different systems should reflect these different goals
    and, therefore, they should be different from a system to another one.

\item $Q_3$. {\em How does the variety of topics a user is concerned with depend on the features and goals of a Social
    Sharing platform?} For instance, do people always focus on the same topics even if they
    shift from a Social Sharing system to another one or do they consider different topics in
    different systems?

\item $H_3$. {\em The variety of user topics varies significantly from a Social Sharing system to
    another one}. We expect that, depending on the features of a Social Sharing system, users tend to state
    different kind of interests and, therefore, they consider different topics in different
    Social Sharing systems.
\end{itemize}

\subsection{Correlations between tagging behavior and friending behavior}

A second research theme we focus on is about the relationships that exist between social and tagging behaviors. In
detail, we aim at answering the following research questions:

\begin{itemize}

\item $Q_4$. {\em Does the intensity of the activity a user carries out in a Social Sharing
    system depend on the type of activity?}. In other words, we are interested in assessing
    whether a form of correlation exists between the frequency of tagging activities of a user
    and the number of her social contacts.

\item $H_4$. {\em The intensity of tagging activity is related to the intensity of friending
    activity and vice versa}. We believe that in each Social Sharing system there is a small
    fraction of {\em active users} who are densely connected with other users and, at the same
    time, generate a large volume of contents/tags in the platform.

\item $Q_5$. {\em Does the variety of tags a user exploits correlate with the number of her
    social contacts?} In other words, is there a relationship between the number of distinct
    tags composing the profile of a user and the number of tags in the profiles of her friends?

\item $H_5$. {\em The variety of tags exploited by a user is related to the number of her
    social contacts}. According to the {\em homophily principle} in sociology (which states that user
    with similar tastes are likely to be friends), we guess that a user share one or more
    interests with her friends. Therefore, users with a great number of friends should also
    present highly diversified interests.

\end{itemize}

\subsection{Methodology}
\label{sub:methodology}

To answer the research questions formulated above, we analyzed user profiles in Social Sharing systems at three
different levels of granularity:

\begin{itemize}

\item {\em Intensity level}. At this level we studied user behaviors on the basis of the
    ``volume'' of the activities they carried out. We started our analysis by
    comparing the {\em frequency of tagging} $ft_u$ of a user $u$ in different social systems, i.e., the number of tags that a user applied. We
    studied also the frequency of social interactions $fs_u$ of $u$ in different social system, i.e., the number of friends of $u$. We compared $ft_u$ and $fs_u$ in different social systems.
\item {\em Tag level}. At this level we focused on the tag-based profile of a user. We started
    studying whether these profiles focused on few (and often frequently used) tags or, vice
    versa, if users prefer to use a large number of tags. In the first case, user interests
    appear strongly polarized towards few tags. Therefore, it is easy to predict, for
    instance, what kind of objects a user will look for in the future. By contrast, if tags are
    applied in a uniform fashion, the amount of randomness in a profile is high and, therefore,
    the task of predicting future user interests is hard. We compared the frequency of tagging by a user with that of
    her friends to study whether significant differences emerge.

\item {\em Semantic level}. In addition to analyzing the raw tags that people use to describe
    resources, the semantic analysis investigates the actual concepts that people refer to. In
    particular, we analyzed to what extent tag-based profiles can be mapped onto {\em (i)}
    {\em lexicographic} and {\em (ii)} {\em ontological} concepts in the context of different Social
    Sharing systems. Moreover, we investigated to what extend the variety of semantic profiles
    depends on the given environment in which a user performed the corresponding tagging
    activities.

\end{itemize}

\section{Datasets}
\label{sec:datasets}

In order to carry out our analysis we collected the public profiles of 421,188 distinct users by
means of the Google Social Graph API. Data were collected from September 2009 to January 2011. Such
an API is capable of finding connections among persons on the Web. It can be queried through an
HTTP request having a URL called {\em node} as its parameter. The node specifies the URL of a Web
page of a user $u$.

The Google Social Graph API is able to return two kinds of results:

\begin{itemize}

\item {\em A list of public URLs that are associated with $u$}; for instance, it reveals the
    URLs of the blog of $u$ and of her Twitter page.

\item {\em A list of publicly declared connections among users}. For instance, it returns the
    list of persons who, in at least one social network, have a link to a page which can be
    associated with $u$.

\end{itemize}

The whole dataset used in this paper is available at the following URL: {\tt http://www.wis.ewi.tudelft.nl/data/crosstags.tar.gz}.

After collecting data, we observed that 1,467 users had a profile on Flickr and Delicious and only 321 users had
a profile in all the three systems.


The tagging statistics of the users at Flickr, Delicious, and
StumbleUpon are listed in Table~\ref{table:folksonomy-stats-fds} (\emph{FDS dataset}). Overall,
these users performed 387,786~tag assignments (TAS). On Flickr, users tagged most actively with, on
average, 532.99 tag assignments, followed by Delicious (483.58 TAS) and StumbleUpon (191.48 TAS).
It is interesting to see that Delicious tags constitute the largest vocabulary although most of the
tagging activities were done in Flickr: the Delicious folksonomy contains 21,239 distinct tags while
the Flickr folksonomy covers 18,240 distinct tags. Correspondingly, tag-based Delicious
profiles contain, on average, 66.17 distinct tags in contrast to 56.82 distinct tags for the Flickr
profiles.

Table~\ref{table:folksonomy-stats-fd} lists the tagging statistics of those users who have a Flickr \emph{and} Delicious account, but are not necessarily registered to StumbleUpon ({\em FD} dataset).

\begin{table}[t]
\begin{center}
\tbl{FDS dataset: Tagging statistics 
for the 321 users who have an account at Flickr, Delicious, and StumbleUpon.\label{table:folksonomy-stats-fds}
}{
\small
\begin{tabular}{|l|c|c|c|c|c|}
\hline
 &  \multirow{2}{*}{\textbf{Flickr}} &  \multirow{2}{*}{\textbf{Delicious}} &\textbf{Stumble}&  \multirow{2}{*}{\textbf{All}} \\
 &   &   &\textbf{Upon}&   \\
\hline
\textbf{distinct} & 18,240 & 21,239  & 8,663  & 39,399   \\
\hline
\textbf{TAS} & 171,092 & 155,230  & 61,464  & 387,786   \\
\hline
\textbf{distinct tags/user (avg./median)} & 56.82 / 7  & 66.17 / 11 & 26.99 / 4  & 122.74 \\
\hline
\textbf{TAS/user} & 532.99 & 483.58  & 191.48  & 1,208.06   \\
\hline
\end{tabular}
}
\end{center}
\end{table}

\begin{table}[t]
\begin{center}
 \tbl{FD dataset: Tagging statistics for the 1467 users who have an account at Flickr and Delicious. \label{table:folksonomy-stats-fd}
}{
\small
\begin{tabular}{|l|c|c|c|c|}
\hline
 &  \multirow{1}{*}{\textbf{Flickr}} &  \multirow{1}{*}{\textbf{Delicious}} &  \multirow{1}{*}{\textbf{All}} \\
\hline
\textbf{distinct tags} & \multirow{1}{*}{72,671} & \multirow{1}{*}{59,275}    & \multirow{1}{*}{119,056}   \\
\hline
\multirow{1}{*}{\textbf{TAS}} & \multirow{1}{*}{892,378} & \multirow{1}{*}{683,665}   & \multirow{1}{*}{1,576,043}   \\
\hline
\textbf{distinct tags/user (avg./median)} & \multirow{1}{*}{49.54 / 51} & \multirow{1}{*}{40.41 / 169}    & \multirow{1}{*}{81.16} \\
\hline
\multirow{1}{*}{\textbf{TAS/user}} & \multirow{1}{*}{608.30}  & \multirow{1}{*}{466.03}  & \multirow{1}{*}{1,074.33}   \\
\hline
\end{tabular}
}
\end{center}
\end{table}

\begin{figure}%
		\includegraphics[width=\columnwidth]{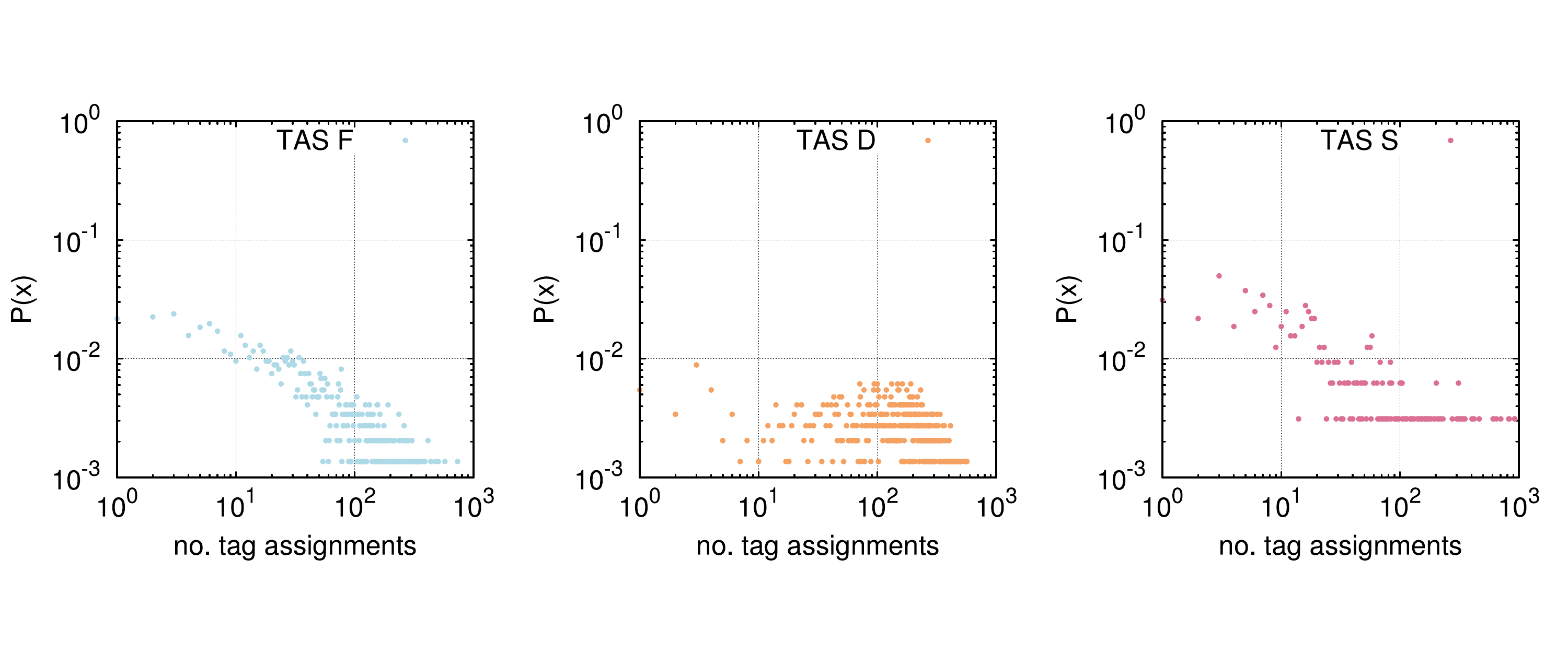}
		\caption{Number of tag assignments performed by the users ({\em FDS} dataset)}%
		\label{fig:tas}%
\end{figure}
\begin{figure}%
		\includegraphics[width=\columnwidth]{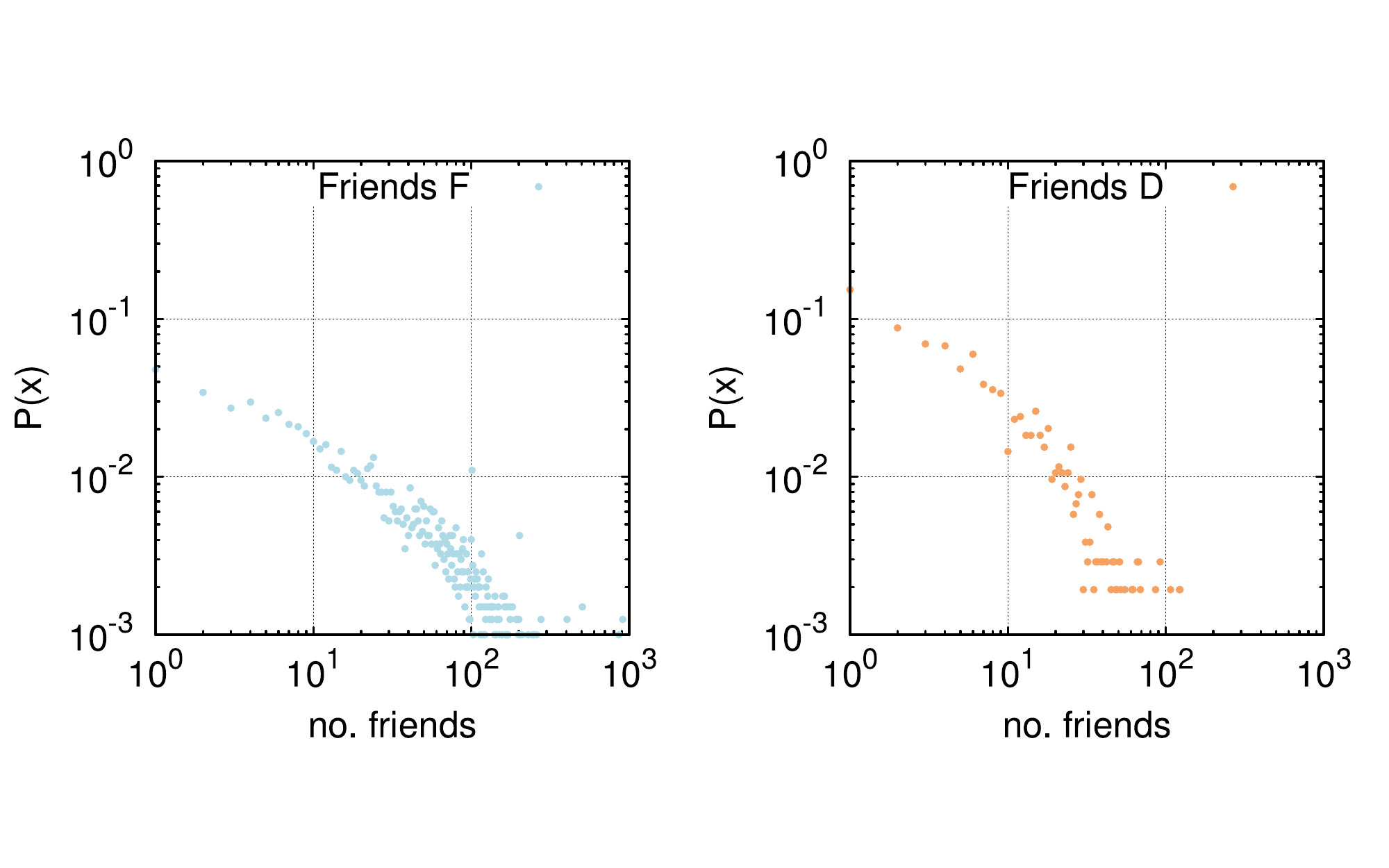}%
		\caption{Number of friends per user ({\em FD} dataset)}%
		\label{fig:friends}%
\end{figure}

Figure~\ref{fig:tas} depicts the number of tag assignments for the users of the {\em FDS} dataset. For the three platforms, we
observe that there is a small fraction of users who performed a large number of tag assignments.
For example, less than 5\% of the users performed more than 500~tag assignments in Flickr and
Delicious respectively. The majority of users is rather moderately active: for
example, in Delicious, approximately 70\% of the people perform between 50 and 300 tag assignments.

As for social behaviors, we have restricted our attention only on friendship relationships. In fact, in Flickr groups have a
great relevance because they offer users the chance of sharing their photos with a worldwide audience and, due to this reason, Flickr groups are often able to attract a large number of users. By contrast, groups in Delicious are less important and their size appear to be comparatively smaller than the typical size of a Flickr group.

As for the number of friends, a heavy-tailed like distribution also emerges (see
Figure~\ref{fig:friends}). For example, there are only a few users (less than 5\%) who have more
than 500~friends on Flickr. On Delicious, people have much less friends but the distribution of
the number of friends of a user is shaped as a heavy-tail.

\section{User Profile Analysis at the Intensity level}
\label{sub:intensity} In this section, we analyze user profiles at the {\em intensity level}. In
detail, we first analyze the {\em deviation of the tagging frequency} as well as the {\em deviation
on the number of friends} for individual users. Furthermore, we investigate whether there are correlations between tagging and friending
activities of a user within the specific Social Sharing platforms.

%

\subsection{Deviation in the frequency of tagging}
\label{subsub:freqtagging}
In this scenario we considered Delicious ($D$), Flickr
($F$) and StumbleUpon ($S$) in a pairwise fashion. For sake of simplicity, let us first consider
Delicious and Flickr together. In this case, we considered only users who applied at least one tag
in both Delicious and Flickr. We define the {\em tagging deviation} $td_u^{DF}$ of a user $u$ as
\begin{equation}
\label{eqn:tagdev}
td_u^{DF} = \frac{|ft_u^D - ft_u^F|}{\max\{ft_u^D,ft_u^F\}}
\end{equation}

Here $ft_u^D$ (resp., $ft_u^F$) is the frequency of tagging of $u$ in Delicious (resp.,
Flickr).

\begin{figure}[t!] \centering
\subfigure[Flickr vs. Delicious]{
 \label{fig:TD-DF}
 \begin{minipage}[tb]{0.3\textwidth}
  \centering
  \includegraphics[width=\textwidth]{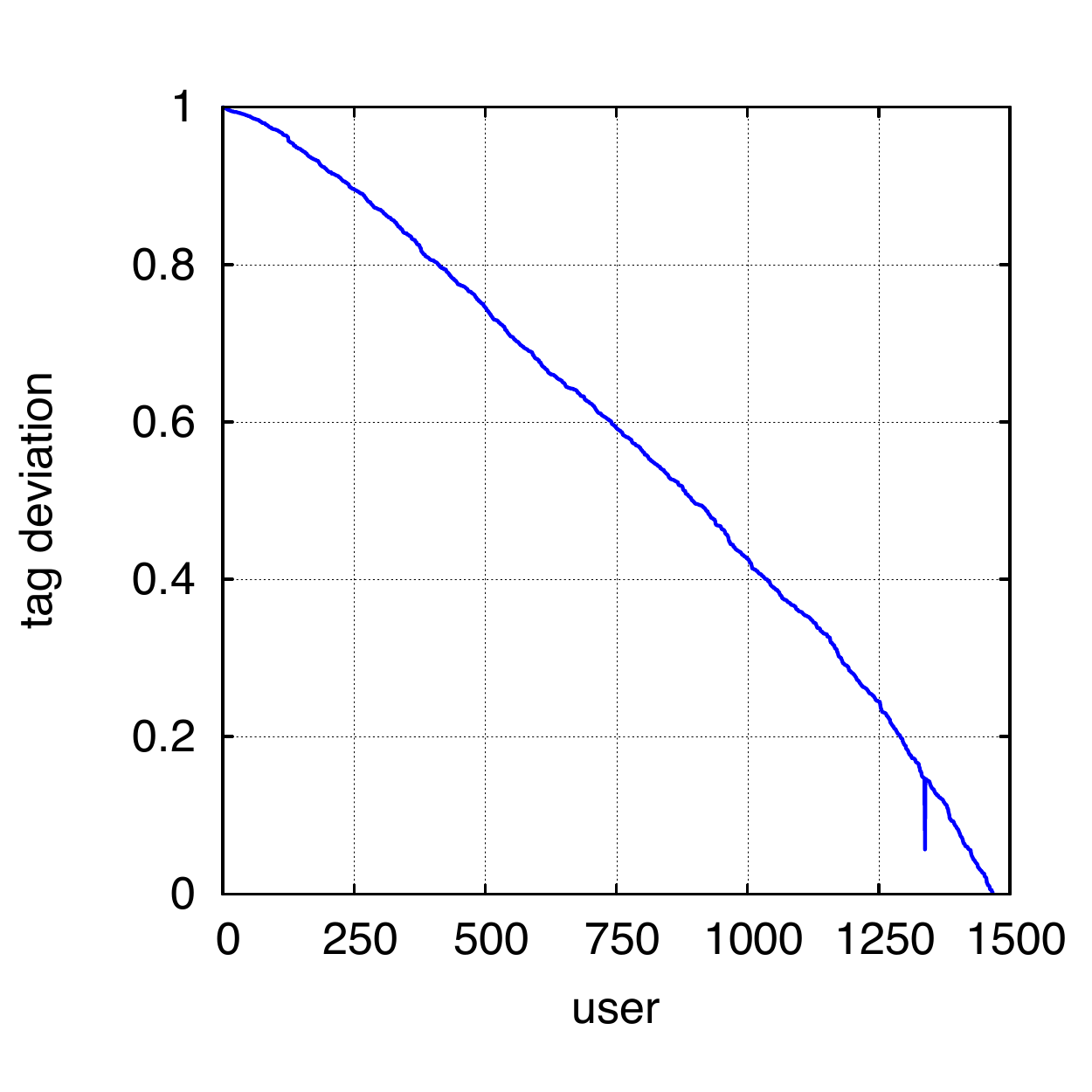}
   \end{minipage}}
\subfigure[Delicious vs. StumbleUpon]{
 \label{fig:TD-DS}
 \begin{minipage}[tb]{0.3\textwidth}
  \centering
  \includegraphics[width=\textwidth]{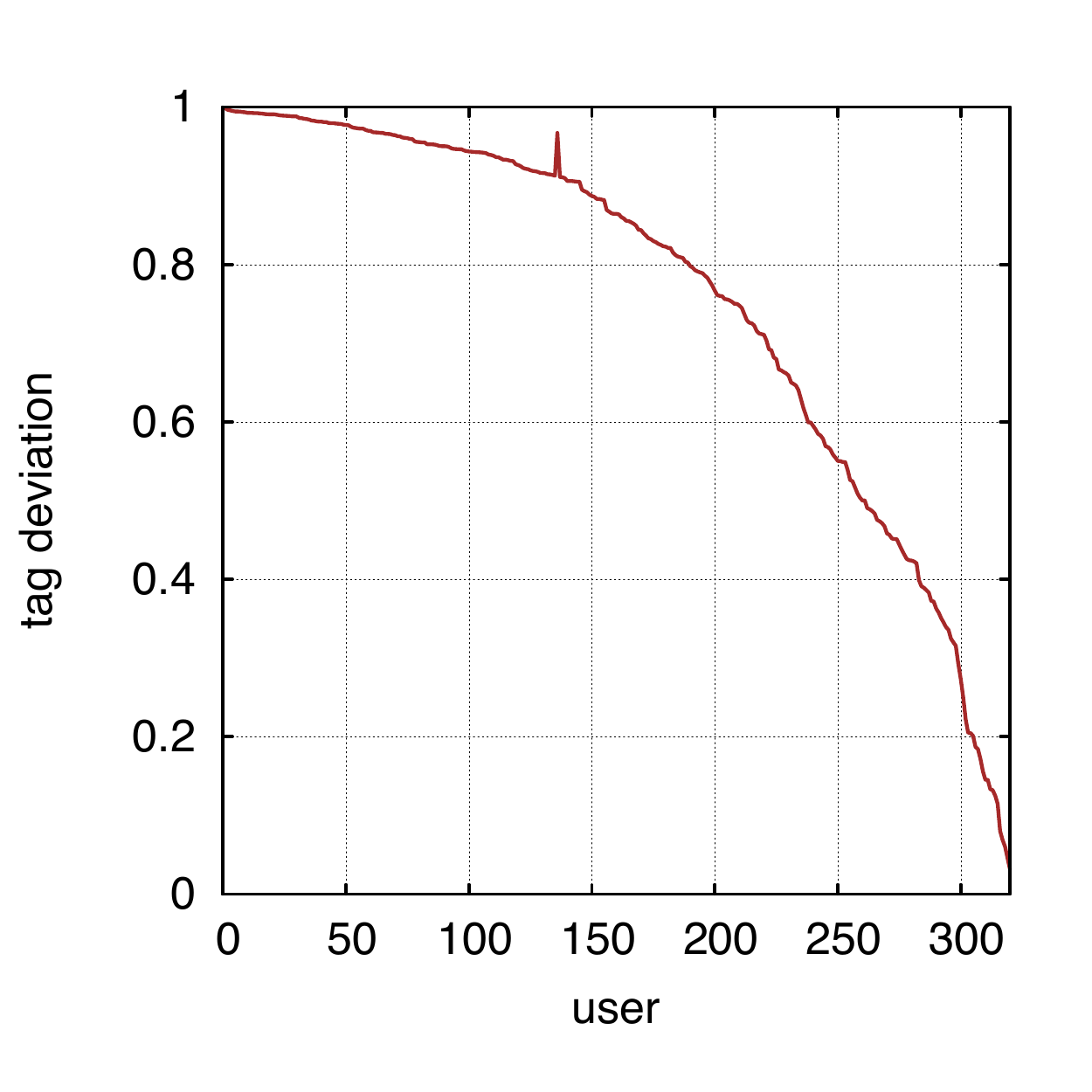}
   \end{minipage}}
\subfigure[Flickr vs. StumbleUpon]{
 \label{fig:TD-FS}
 \begin{minipage}[tb]{0.3\textwidth}
  \centering
  \includegraphics[width=\textwidth]{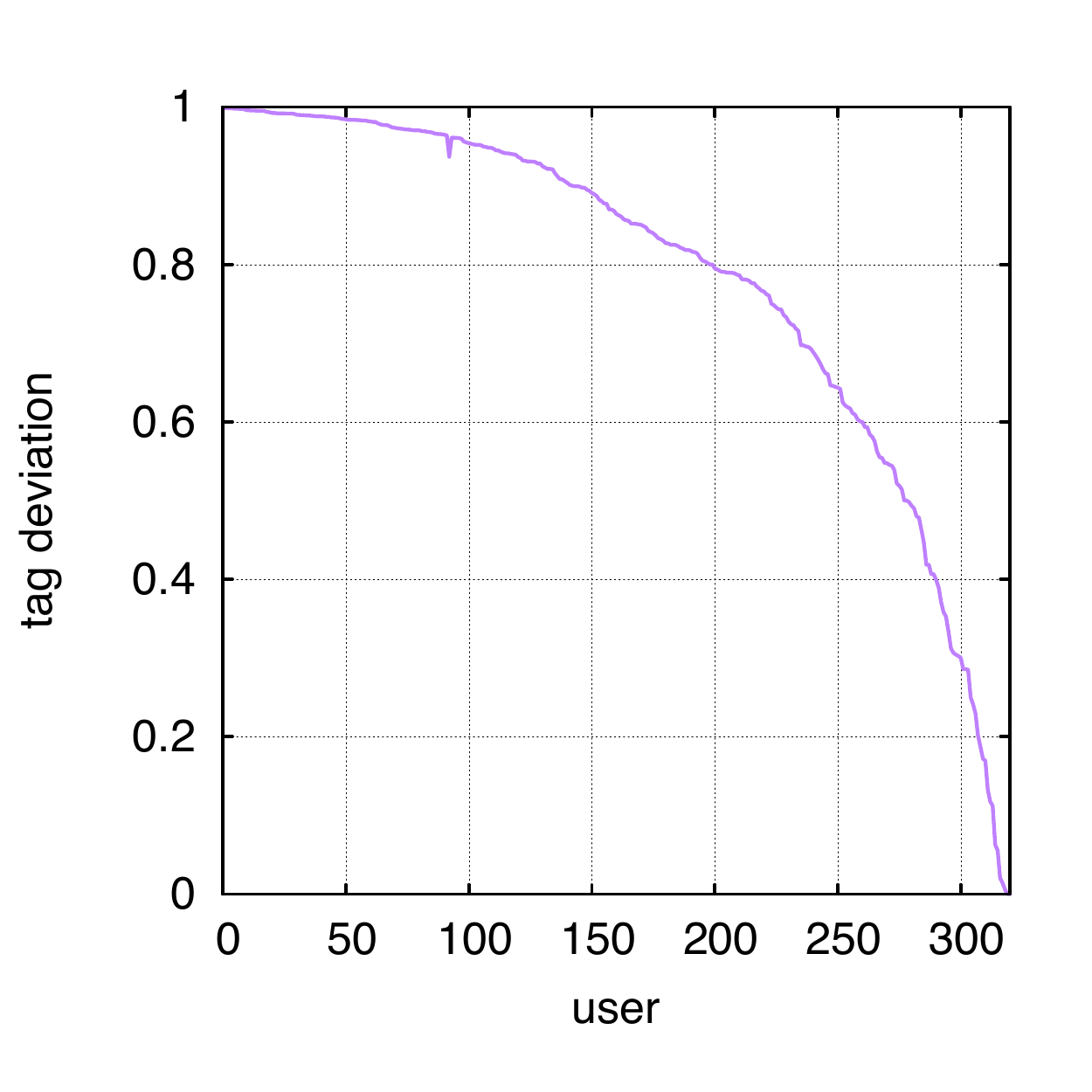}
   \end{minipage}}
 \vspace{-0.2cm}
\caption{Tag Deviations of users tag-based profiles within the different social tagging environments.}
\label{fig:tag-deviation}
\end{figure}

The parameter $td_u^{DF}$ ranges in [0,1]: if $td_u^{DF}$ approaches 0, $u$ tends to approximatively use the same number of tags both in Delicious and in Flickr.
In an analogous fashion, we can define $td_u^{DS}$ and $td_u^{FS}$ to compare StumbleUpon, respectively with Delicious and Flickr.
We sorted $td_u^{DF}$, $td_u^{DS}$ and $td_u^{FS}$ in decreasing order and plotted them in Figures \ref{fig:TD-DF},
\ref{fig:TD-DS} and \ref{fig:TD-FS}.

From Figure \ref{fig:TD-DF} we can observe that $td_u^{DF}$ decreases in an almost linear fashion. This means that the number of users showing ``high'' values of $td_u^{DF}$ is roughly the same as the number of users who show ``low'' values of $td_u^{DF}$.

The first consequence is that in Delicious and Flickr, there is no {\em dominant
behavior} and that the number of users who frequently apply tags in both of the two Social Sharing
systems is roughly equal to the number of users who prefer to focus their tagging activities in
just one of them.

Strong differences, instead, emerge from Figures \ref{fig:TD-DS} and \ref{fig:TD-FS}. In detail, as
for Delicious and StumbleUpon, only about 10\% of users show
a tag deviation less than 0.4. This means that their tagging activities are more {\em
focused}. In general, users who heavily tag in a Social Sharing system do
not tag with the same frequency in the other one. An analogous behavior emerges for Flickr and
StumbleUpon. Hence, since there are many users for which the tag deviation is close to~1 and few
users for whom the tag deviation is close to~0, we conclude that the user behavior varies stronger between Delicious and StumbleUpon
(Figure~\ref{fig:TD-DS}) and Flickr and StumbleUpon (Figure~\ref{fig:TD-FS})
than between Flickr and Delicious (Figure~\ref{fig:TD-DF}). 

\begin{figure}%
	\centering	
		\includegraphics[width=0.45\textwidth]{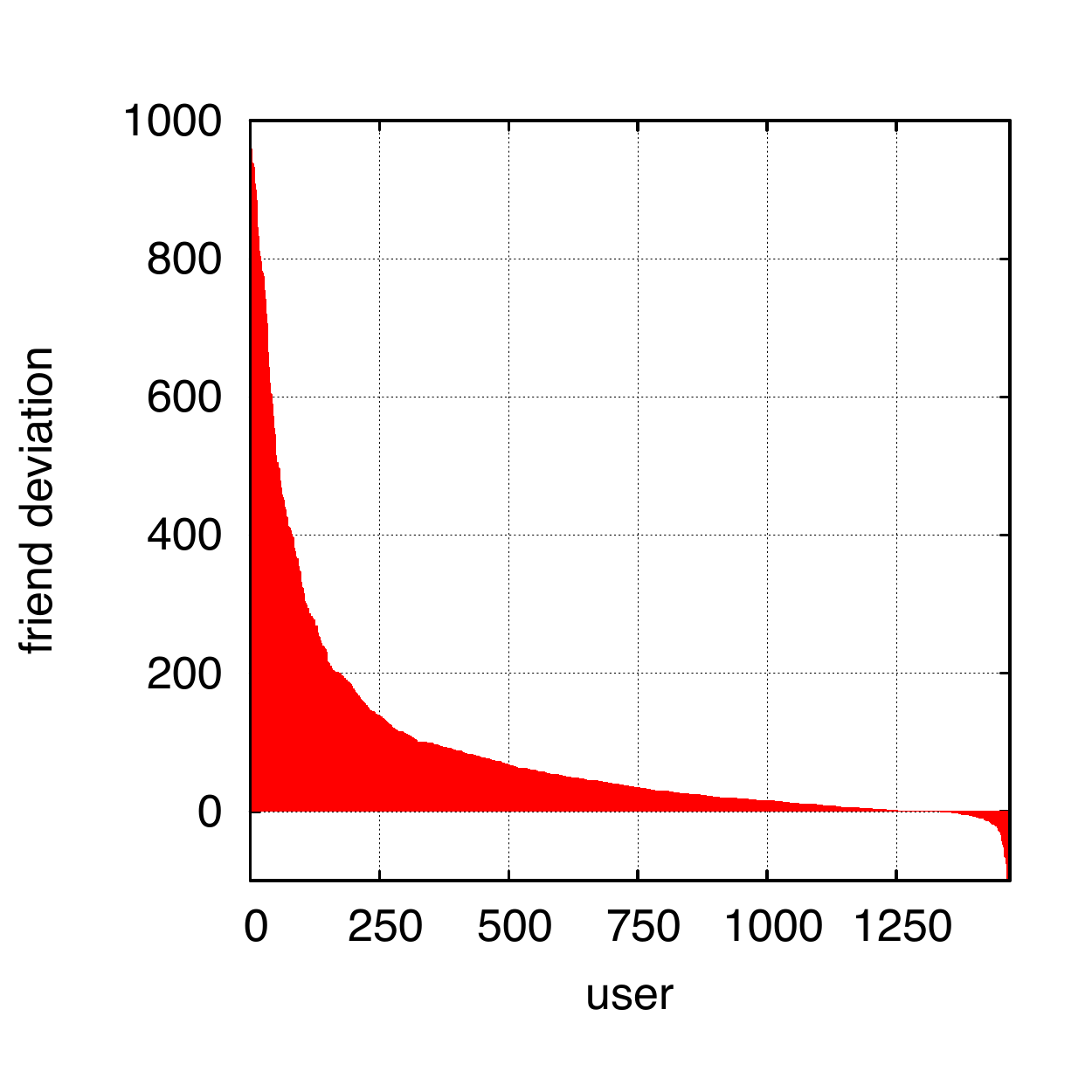}%
		\caption{Friend Deviation (Delicious and Flickr)}%
		\label{fig:FD-FD}%
\end{figure}

The differences in tagging behavior (and, in particular, in the frequency of tagging) can be
explained by taking into account the ultimate goal of each system.

In particular, in Delicious and Flickr, users are prompted to apply tags for quickly retrieving of contents of their interest or to share a content they deem as relevant with other users.
Therefore, the higher the number of exploited tags, the easier is to retrieve relevant resources.
In StumbleUpon, vice versa, the need of tagging resources is less urgent; StumbleUpon, in fact, relies on human opinions which are expressed by reviews or by applying positive/negative feedbacks which are combined through Collaborative Filtering algorithms.
Therefore, the ultimate goal of users in StumbleUpon is finding novel content rather than labeling contents to easily retrieve
them.
This explains the lower level of tagging activity of StumbleUpon users.

\subsection{Deviation in the frequency of social interactions}
\label{subsub:freqsocial}
In an analogous fashion, we studied the quantitative difference in users behavior according to the frequency of social interactions.
As pointed out in Section \ref{sub:social-systems-description}, we focused only on Delicious and Flickr because StumbleUpon does not support social contacts among its users.
Given a user $u$, the {\em friend deviation} $fd_u^{FD}$ is defined as
\begin{equation}
\label{eqn:frienddeviation}
fd_u^{FD} = fs_u^{F} - fs_u^{D}
\end{equation}

being $fs_u^{F}$ (resp., $fs_u^{D}$) the number of friends of $u$ in Flickr (resp., Delicious).
Note that $fd_u^{FD}$ can feature both positive values (if $fs_u^{F} > fs_u^{D}$) and negative ones (in the opposite case). We sorted users on the basis of decreasing
values of $fd_u^{FD}$ and plotted the corresponding values in Figure~\ref{fig:FD-FD}.

We may observe that $fd_u^{FD}$ is positive for about 1300 users
(out of 1467). This means that users in our dataset usually prefer to create friendships in Flickr
rather than in Delicious. The architecture of Flickr, therefore, creates an environment in
which social networking activities are---in addition to tagging activities---essential for the
utility of the system itself.
The architecture of Delicious, on the other hand, seems to lack corresponding features that create
a social networking environment that people enjoy to utilize.

The value of $fd_u^{FD}$ is larger than 200 for about
150 users (out of 1467), i.e., there are users who have, in Flickr, more than 200 friends than
those they have in Delicious. We believe that most of these relationships are not to be intended as
{\em true friendships} and the actual number of friends of a user is much smaller. Our conjecture
relies upon some studies in anthropology that reveal that there is an upper limit to the number of
people with whom a user can maintain stable social relationships, and, therefore, that can be
considered as {\em actual friends}. This number is known as {\em Dunbar number}
\cite{dunbar1998grooming} and its value is around 130. Recently, some authors computed the average
number of friendship relationships in online social network like  Facebook~\cite{ferrara2011crawling,ferrara2011community,ferrara2012large} reporting that, in this context, the average number of friends of a user is between 100 and 150, in agreement with our findings.

\subsection{Correlation between the frequency of tagging and the frequency of social interactions}
\label{subsub:socialtagging}
We studied the relationship between the frequency of tagging $ft_u$ of a user $u$ and her frequency of social interactions $fs_u$.
We focus again on Flickr and Delicious because StumbleUpon does not feature social connections.

To graphically analyze this correlation, we used a bi-dimensional plane such that each user $u$ is
mapped onto a point in this plane. The abscissa of this point represents $ft_u$ whereas the
ordinate coincides with $fs_u$. The obtained results are reported in Figure~\ref{fig:friends-tag-flickr}
for Flickr and in Figure~\ref{fig:friends-tag-delicious} for Delicious.
From the analysis of these figures, we may observe that the plane can be split into four regions:

\begin{figure}%
	\centering
	\begin{minipage}{.49\columnwidth}
		\includegraphics[width=\columnwidth]{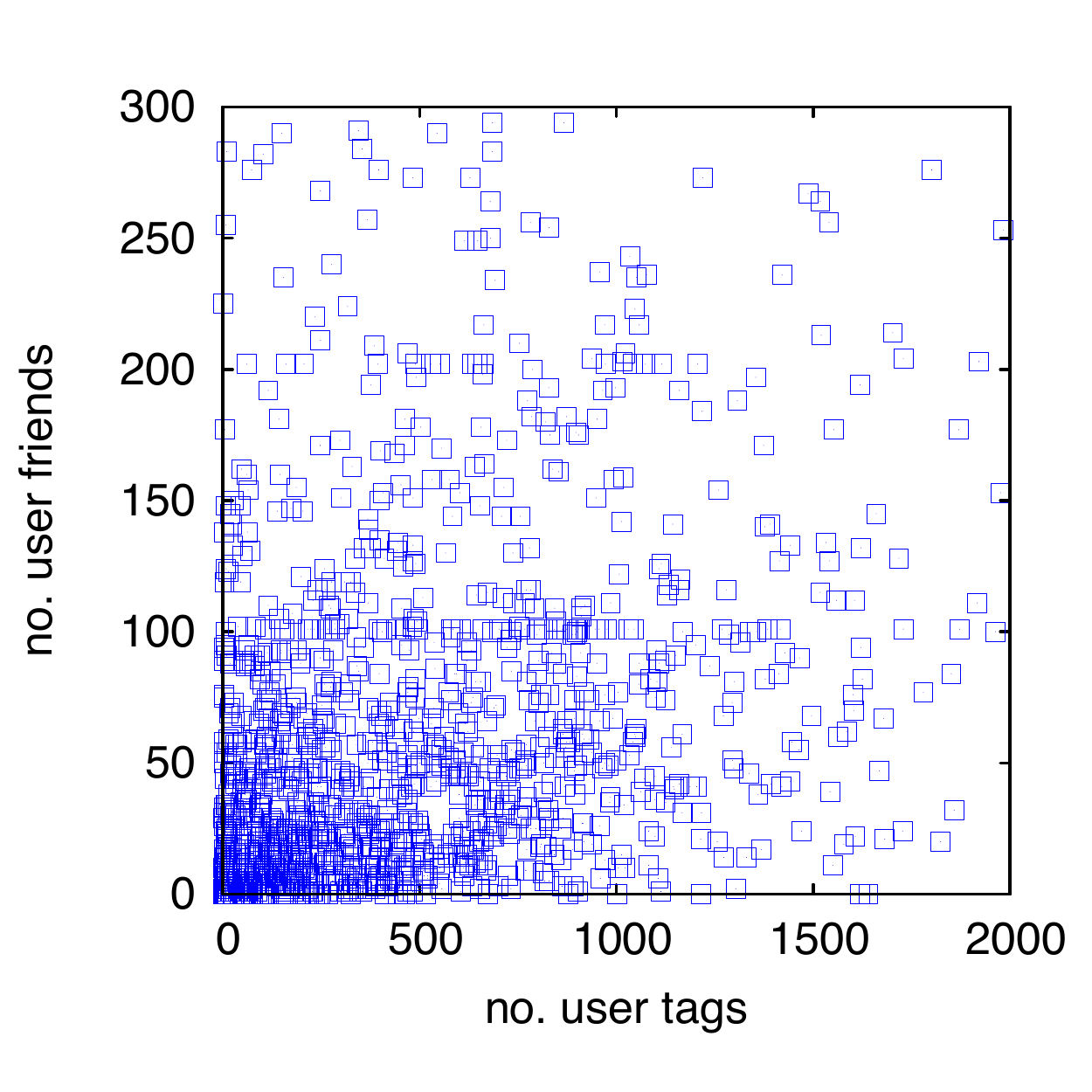}%
		\caption{Correlation between $fd$ and $fs$ in Flickr}%
		\label{fig:friends-tag-flickr}%
	\end{minipage}
	\begin{minipage}{.49\columnwidth}
		\includegraphics[width=\columnwidth]{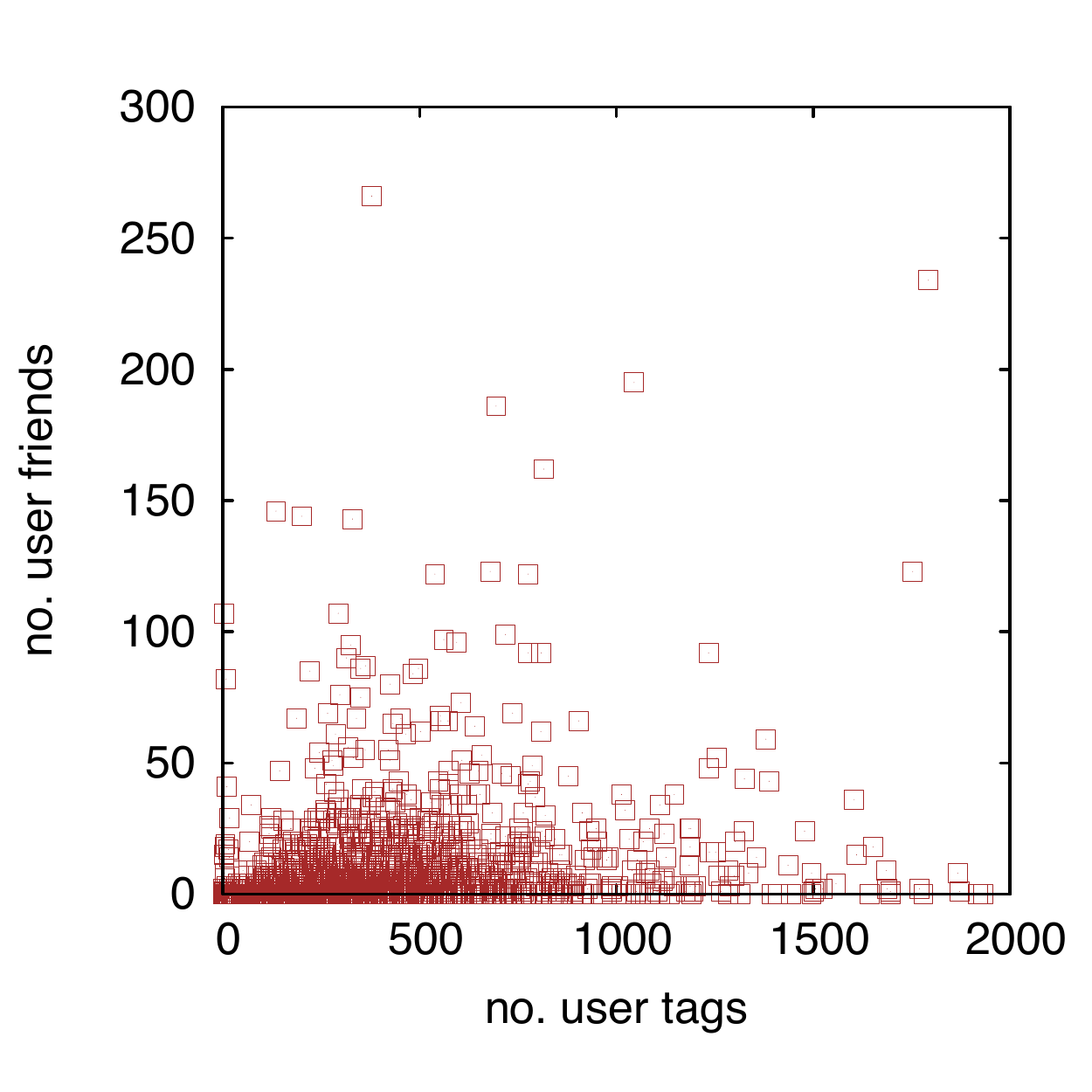}%
		\caption{Correlation between $fd$ and $fs$ in Delicious}%
		\label{fig:friends-tag-delicious}%
	\end{minipage}	
\end{figure}

\begin{enumerate}

\item There is a dense region consisting of users who have few friends and, {\em at the same
    time}, have applied only a few tags. Such a region emerges in both Flickr and Delicious (SW quadrant). In
    detail, an inspection of the diagrams reported in Figures \ref{fig:friends-tag-flickr} and
    \ref{fig:friends-tag-delicious} indicates that most of the examined users have less
    than 150 friends and applied less than 1,000 tags in Flickr; as for Delicious, most of the existing
    users in our dataset have less than 50 friends and exploited less than 1,000 tags as well.

    The existence of such a region depends on two facts which {\em jointly co-occur}: {\em (a)}
    The number of friends of a user follows a heavy-tailed distribution and, therefore, the typical number of
    friends of a given user is moderately small. {\em (b)} The number of tags applied by a user follows a
    heavy-tailed distribution and, therefore, the typical number of tags applied by a given user is quite small.

\item There is a region consisting of users who applied a large number of tags and {\em at the
    same time}, they have a large number of friends (NE quadrant). The user population density in such a
    region is extremely low both for Flickr and Delicious and, again, it depends on the broad distributions
    of tag usage and friendship (cf. Figure~\ref{fig:tas}
    and Figure~\ref{fig:friends}).

\item There is a third region formed by users with a large number of friendships but
    who used comparatively few tags than other users (NW quadrant). The user
    population density in this region in Flickr was higher than in Delicious. Such a behavior
    is likely to depend on the fact that in Flickr the tendency to socialize becomes more
    marked than in Delicious; as a consequence, there exists a significant portion of Flickr
    users who intend Flickr as a platform to promote and foster the creation of social
    relationships rather than a mere social tagging platform.

\item There is a fourth region populated by users who apply a large number of tags but, at the
    same time, have few friends (SE quadrant). Such a region is more dense in Delicious than in Flickr and it
    highlights an opposite behavior against to that discussed in Point~(3). In detail,
    in Delicious, tagging is the main user activity; by contrast, the friending activities oriented to ``socialization'' acquire a less
    relevant role.
\end{enumerate}

\subsection{Findings}
\label{subsub:findings1}
Given the results from our analysis at the intensity level of tagging and friending activities,
we can summarize our main findings as follows:

\begin{enumerate}

\item As for tagging activities, users who are highly active in Delicious may also be highly active in
    Flickr. Moreover, there is a large fraction of users that is highly active in either one of the two
    environments. Therefore, the aptitude of a user to apply tags does not seem to depend on the design of these
    two systems. Such a conclusion, however, does not hold when comparing the
    users' profiles from StumbleUpon with the ones from Flickr or Delicious. Here, we observe that
    the tagging design has a significant impact on the tagging behavior of the users.

\item As for social activities, Flickr is designed to serve as a social network because users tend to creating and cultivating social
    relationships. Such a behavior does not emerge in Delicious: the number
    of friendships of a user in Delicious is, on average, smaller than the number of her friends in Flickr. Therefore, the tendency of a user to socialize depends on the system in
    which she operates in.

\item For Flickr, we observe that there is a correlation between the intensity of the tagging
    activity and the intensity of the friending activities. In particular, a low
    tagging activity often implies a low number of social interactions (and vice versa)
    while users who intensively tag show also a solid intensity
    regarding their friending activities. However, in
    Delicious, we observe that the intensity of tagging activities tends to be larger than the
    intensity of friending for the majority of the users.

\end{enumerate}

A possible explanation to these findings is that in Flickr, users are incentivized to apply
tags and create friendships relationships because the platform represents a stimulant environment to popularize their photos.
By means of tags and social relationships users can share with others important facts of their life
(e.g., graduation party) as well as they can discover, in a serendipitous fashion, novel photos.
Such a kind of incentive is less important in Delicious/StumbleUpon because users often tend to use
these platforms as a repository of resources they deem as important and there
is no urgent need to share them with others.

\section{User Profile Analysis at the tag level}
\label{sub:userprofiletagvariety} In this section, we investigate the characteristics of the
information that is featured by the tag-based profiles in the different Social Sharing
environments. We thus consider a tag as a single \emph{information unit} that characterize a user.

\subsection{Entropy of Tag-based User Profiles}
\label{sec:analysis-entropy} We first study and assess the information carried out by the tag-based
user profile $P_T(u)$ of a user $u$ in different Social Sharing systems. In detail, we studied whether $u$ prefers to: {\em (a)} frequently
use few tags or {\em (b)} use a wider set of tags which are applied in a uniform fashion.

To measure the value of the information embodied in tag-based profiles and quantify the
predictability of a user's tagging behavior, we adopt {\em entropy
measure}~\cite{entropyShanon/InformationTheory/1948}. The entropy of a tag-based profile $P_T(u)$ is defined as

\begin{equation}\label{eq:entropy}
entropy(P_T(u)) = \displaystyle\sum_{t \in T_u} p(t) \cdot \left(-\log_2(p(t))\right)
\end{equation}

Here, $T_u$ is the set of tags in the profile of $u$ (see Definition
\ref{def:user-profile}). In Equation~\ref{eq:entropy}, $p(t)$ is the probability that the tag
$t$ was utilized by $u$ and $-\log_2(p(t))$ is called \emph{self-information}.
Using base~2 for the computation of the logarithm allows for measuring self-information as well as
entropy in bits. In order to compute $p(t)$, we used the individual usage frequencies of the tags as follows:

\begin{equation}\label{eq:entropy-tag-prop}
p(t) = \frac{f_u(t)}{\sum_{t_i \in T_u} f_u(t_i)}
\end{equation}

In the formula above, $f_u(t_i)$ is the number of times $u$ applied $t_i$. The higher the entropy of a
tag-based profile, the more valuable the information in the profile. For example, profiles that
feature just one single tag~$\overline{t}$ have an entropy of $0$, because $p(\overline{t})$ would be equal to~1 which
would imply $\log_2(p(\overline{t})) = 0$.

The entropy of a tag-based profile depends on both the size of the profile (number of
distinct tags) and the probability distribution of the profile tags (distribution of usage
frequencies).

Profiles that provide many different tags $t_i \in T_u$ which occur with roughly equal probabilities
$p(t_i)$ have high entropy. These profiles imply a high degree of randomness and, as we
previously pointed out, it would be difficult to predict the tagging behavior of the user.
By contrast, those profiles which contain many different tags but feature only very few tags with
a very high probability (while most tags have a very low probability) will have lower entropy.

\begin{figure}[t!]
\begin{minipage}[tb]{\textwidth}
\end{minipage}
\subfigure[Flickr and Delicious]{
 \label{fig:entropy-fd}
 \begin{minipage}[tb]{0.3\textwidth}
  \centering
  \includegraphics[width=\textwidth]{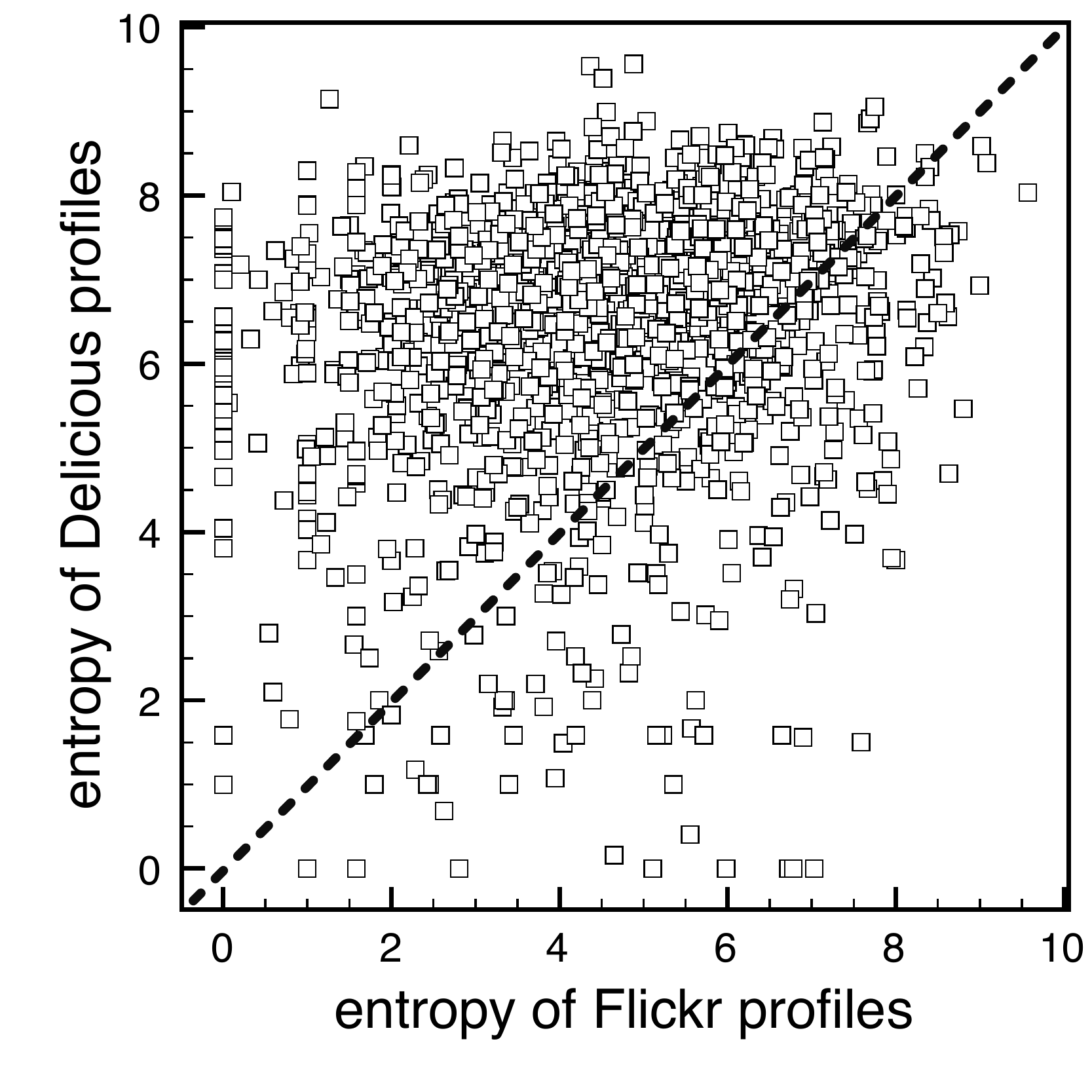}
   \end{minipage}}
\subfigure[Flickr and StumbleUpon]{
 \label{fig:entropy-fs}
 \begin{minipage}[tb]{0.3\textwidth}
  \centering
  \includegraphics[width=\textwidth]{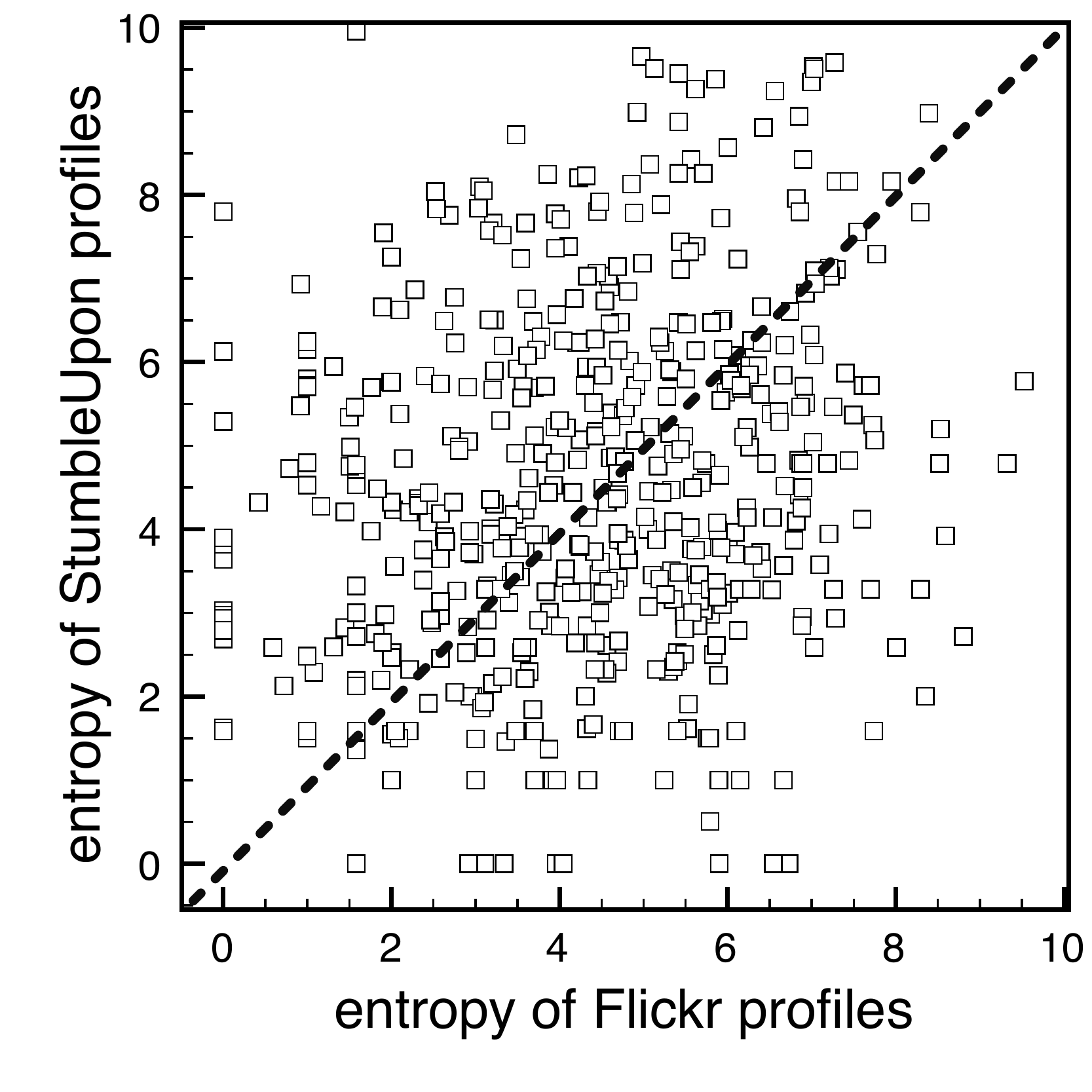}
   \end{minipage}}
\subfigure[StumbleUpon and Delicious]{
 \label{fig:entropy-sd}
 \begin{minipage}[tb]{0.3\textwidth}
  \centering
  \includegraphics[width=\textwidth]{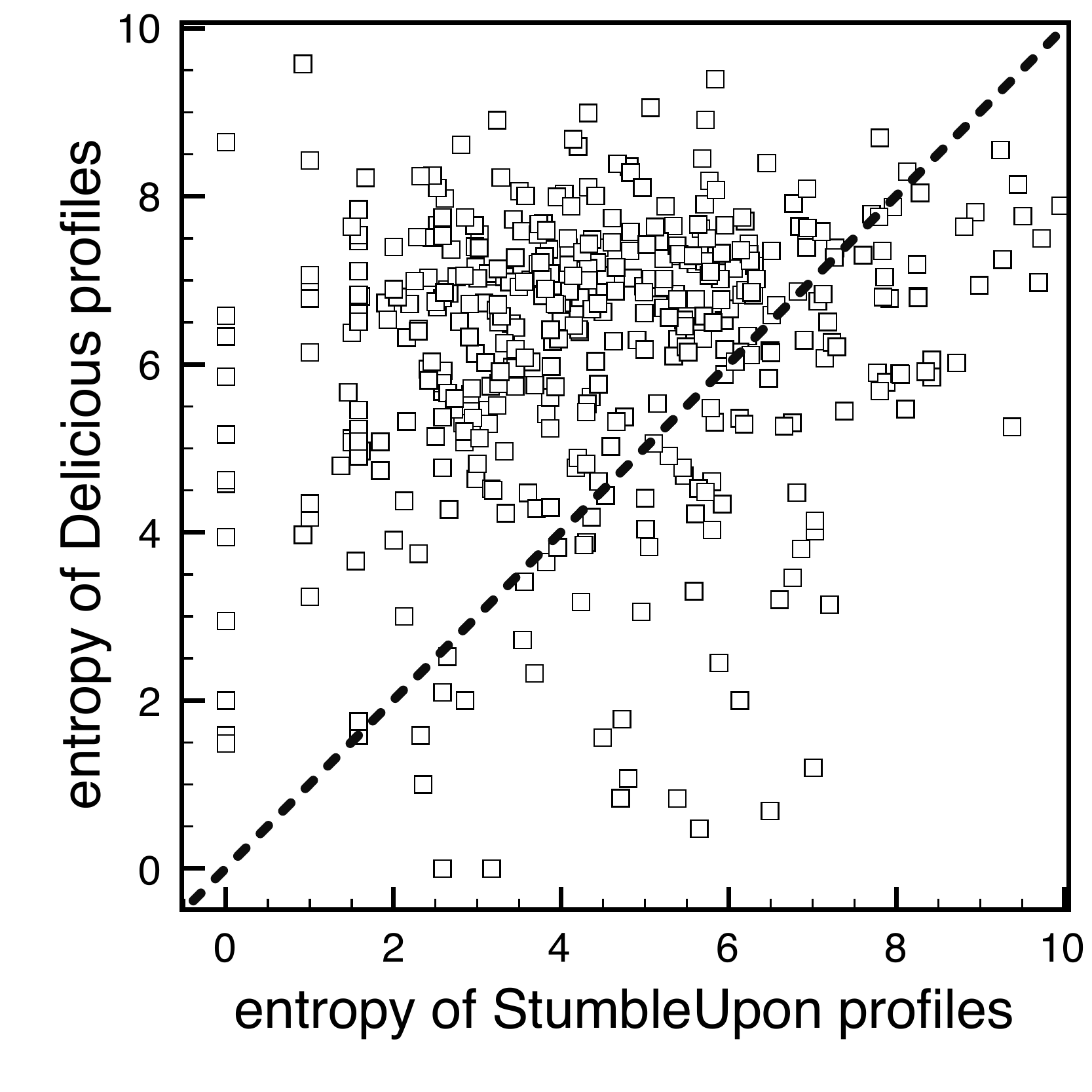}
   \end{minipage}}
 \vspace{-0.2cm}
\caption{Entropy of tag-based profiles (in bits). Each dot represents a user and therefore specifies the entropy of
the user's profile in the two given Social Sharing systems.}
\label{fig:entropy}
\end{figure}

The entropies of user profiles in Delicious, Flickr and StumbleUpon are plotted in Figure~\ref{fig:entropy}: each dot represents a user whose coordinates represent the entropy of her profile
in the two Social Sharing systems reported on the $x$ and $y$ axes, respectively.
From Figure~\ref{fig:entropy} we draw the following conclusions:

\begin{itemize}

  \item For most of the users (81.8\%: those points in Figure~\ref{fig:entropy-fd} which are above the dotted diagonal line; 76.5\% for Figure~\ref{fig:entropy-sd}), the entropy of the Delicious profile is higher than the entropy of the Flickr profile and StumbleUpon profile respectively.
  Given that the size of the Flickr profiles is, on average, higher than the size of the Delicious profiles, it seems that the actual tag usage in Delicious is more heterogeneous than that on Flickr and is therefore also harder to predict.

  \item Flickr and StumbleUpon profiles feature, on average, a similar level of entropy: 4.41~bits for Flickr and 4.59~bits for StumbleUpon.
  Overall, the low entropy of Flickr profiles is an interesting and unexpected characteristic.
  Given that StumbleUpon provides category-like tag suggestions while Flickr allows users to freely choose tags without any tagging support, one would expect that the tagging behavior on StumbleUpon is more predictable than the tagging behavior on Flickr. However, we reveal that the opposite behavior emerges.

  \item On average, the entropy of the Delicious profiles is 6.41 bits. As depicted in Figure~\ref{fig:entropy-fd}, low/high entropy of Delicious profiles does not necessarily imply low/high entropy of Flickr profiles.
  For example, for those users whose Delicious profile has an entropy higher than 6 bits, the entropy of the Flickr profile ranges between 0 and 10 bits.
  Hence, the entropy of the tag-based profiles and the predictability of the tagging behavior is rather influenced by the characteristics of the tagging than by the preferences or characteristics of the individual users.

\end{itemize}

\subsection{Variance of Tag usage frequency distributions}
\label{sub:tagvarianceusage}
To further understand the tag usage of individual users, we analyze the variance of the tag distribution for each user.
This allows us to understand whether users apply tags in a uniform fashion or, vice versa, if the fraction of
tags which are exploited much more (or much less) frequently than others can be regarded as negligible.

More formally, let $T_u$ be the tags applied by $u$ and let $\mu_u$ be the average usage frequency of $u$. The variance $\sigma_u$ is defined as
\begin{equation}\label{eq:variance}
\sigma_u = \frac{\sum_{t_i \in T_u} \left( f_u(t_i) - \mu_u \right)^2 }{|T_u| -1}
\end{equation}

If $\sigma_u\simeq 0$, then $u$ tends to apply tags in a uniform fashion; vice versa, if $\sigma_u$ achieves high values then there are some tags which are much more (resp., much less) frequently applied than others.

\begin{figure}[t!] \centering
\begin{minipage}[tb]{\textwidth}
\end{minipage}
\subfigure[Flickr]{
 \label{fig:variance-flickr}
 \begin{minipage}[tb]{0.3\textwidth}
  \centering
  \includegraphics[width=\textwidth]{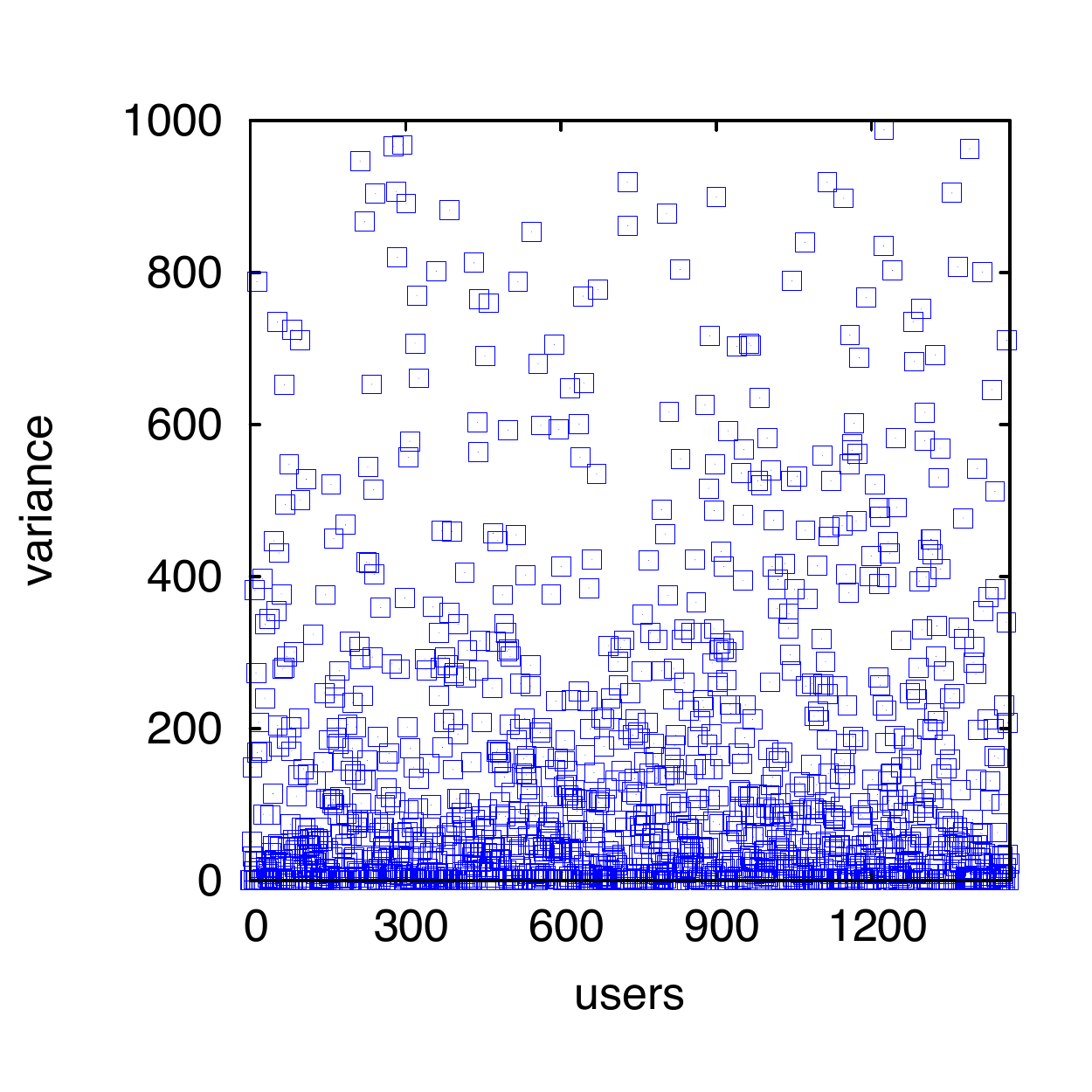}
   \end{minipage}}
\subfigure[Delicious]{
 \label{fig:variance-delicious}
 \begin{minipage}[tb]{0.3\textwidth}
  \centering
  \includegraphics[width=\textwidth]{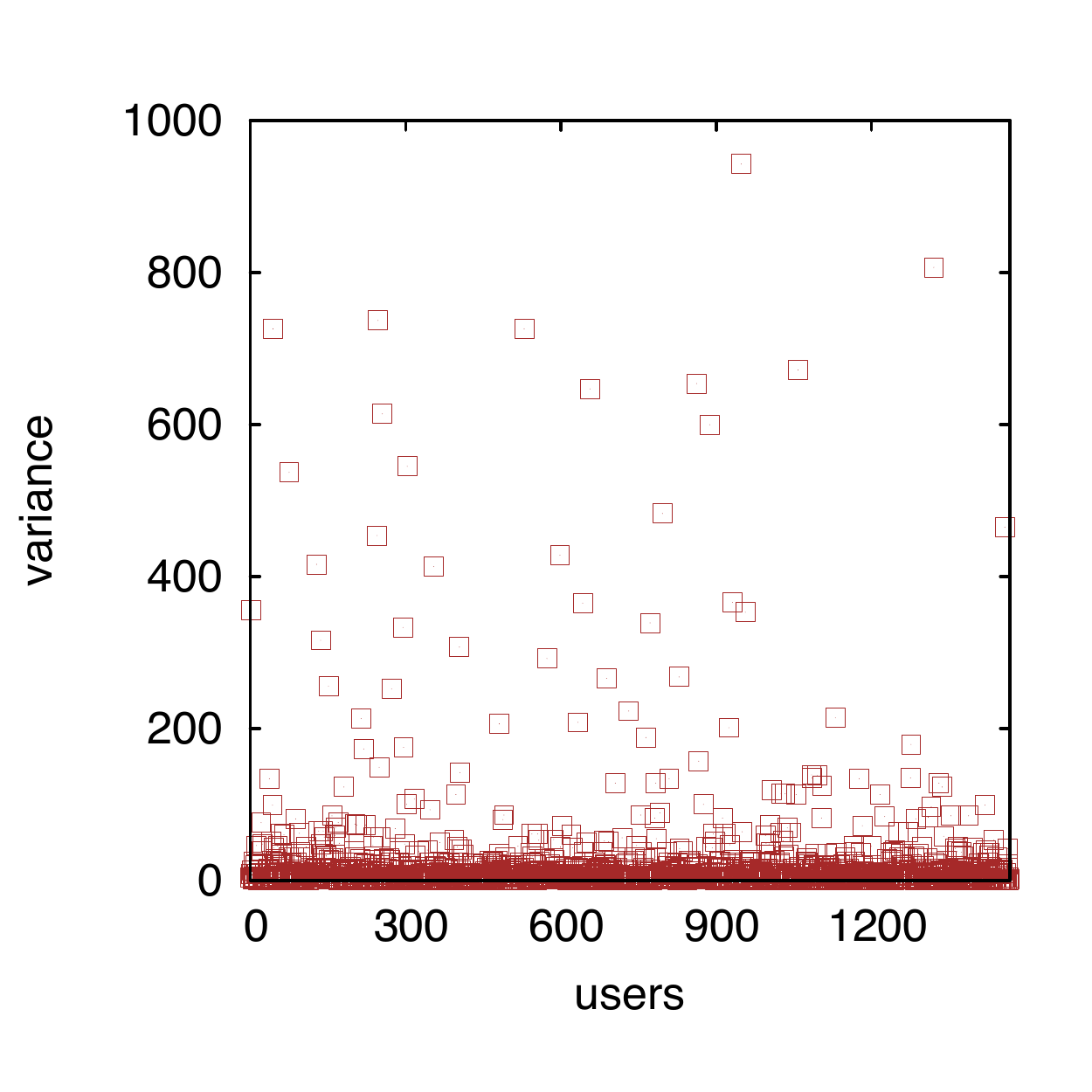}
   \end{minipage}}
\subfigure[StumbleUpon]{
 \label{fig:variance-stumbleupon}
 \begin{minipage}[tb]{0.3\textwidth}
  \centering
  \includegraphics[width=\textwidth]{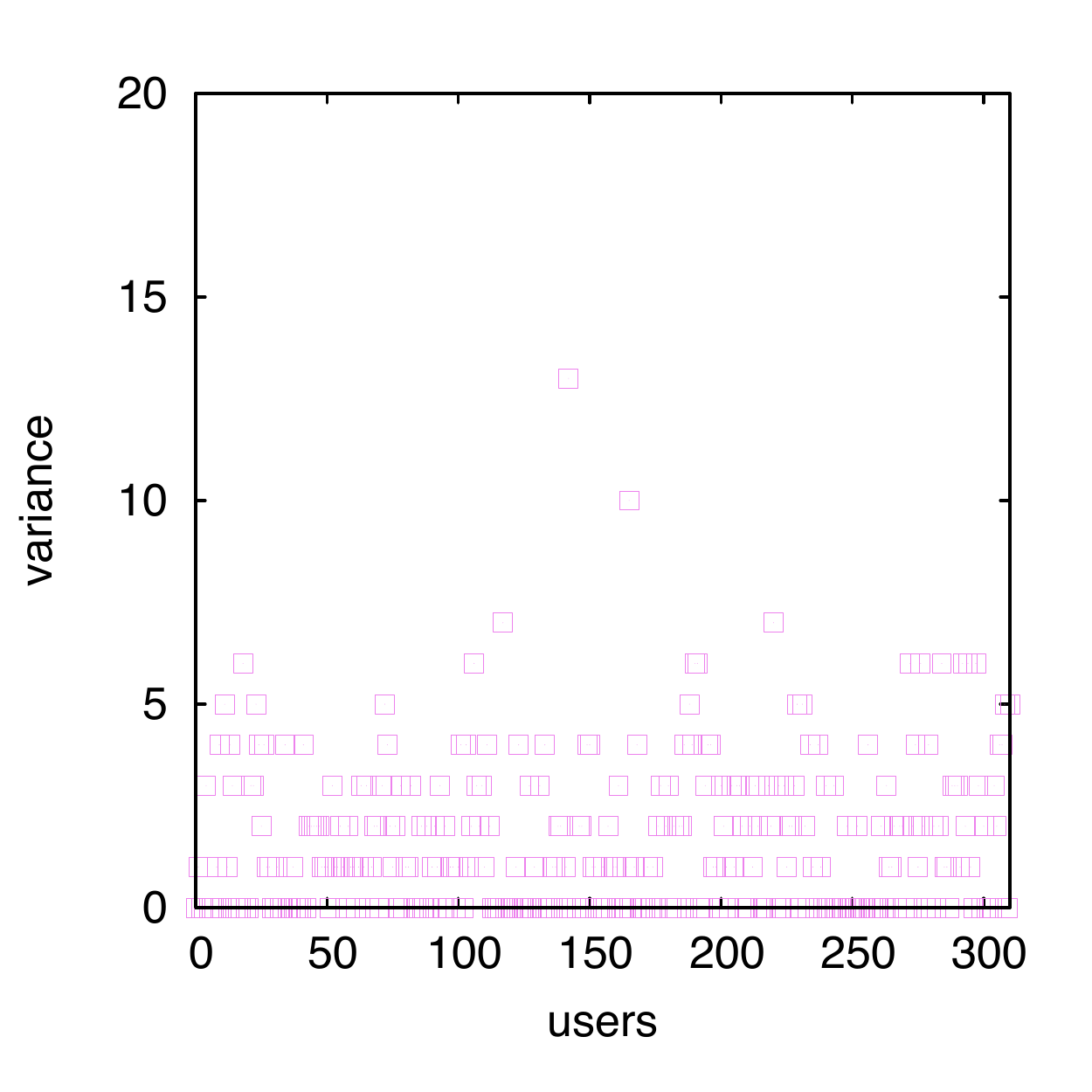}
   \end{minipage}}
 \vspace{-0.2cm}
\caption{Variance of tag usage within the different systems.}
\label{fig:variance}%
\end{figure}

In Figure~\ref{fig:variance}, we plot the variance of the users' tag based profiles in Flickr, Delicious and StumbleUpon respectively.
The highest  occurs for Flickr users, followed by Delicious and StumbleUpon.
This observation is in line with the previous findings where we assessed that Flickr profiles feature, on average, the lowest entropy.

We furthermore compute the {\em largest variance} $\sigma_{max}$, i.e.: $\sigma_{max} = \max_{u} \sigma_u$.
Then, we deduce the percentage $P(k)$ of users who have a variance less than $k \cdot \sigma_{max}$.
Here, $k$ is a threshold value which we set to $k = 0.001, 0.01, 0.05, 0.1$.
Such an analysis is motivated by the following observation: if $P(k) \rightarrow 1$, almost all users
present a variance less than $k\cdot \sigma_{max}$ and, therefore, there is a large mass of users who
tend to apply tags with the same frequency. The obtained results are reported in Table \ref{tab:vardev}.

We observe that 97.61\% of the Delicious user profiles
and 94.61\% of the Flickr user profiles feature a variance which deviates less than 10\% from the
highest value. This means that the vast majority of users tend to apply tags in an homogeneous
fashion. However, as depicted in
Figure~\ref{fig:variance} (and particularly for Flickr), there are also tags (in some user profiles)
which are used much more frequently than others ({\em outlier tags}). Outlier tags cause
a relevant increase of variance scores and they are likely to
reflect personal events in the lives of users and provide a small amount of
information. For instance, the most popular tag in Flickr is {\tt 2010} (it appears 51,726 times)
and other popular tags are {\tt vacation} (43,993 times), {\tt family} (41,995 times) and {\tt
wedding} (37,381 times).

The variance plot for Delicious and Flickr resembles a cloud of
points whereas the StumbleUpon plot is a set of stripes of points. This is due to
the specific goal of tagging in StumbleUpon: in fact, in Delicious or Flickr users are allowed to
freely type tags or select a tag among those provided by other users and recommended by the system. StumbleUpon, by contrast,
provides users with suggestions about categories of tags to use: so, for instance, while in
Delicious a user is allowed to apply tags like {\tt Java} or {\tt C++} (which are regarded as
independent), in StumbleUpon there is a unifying category of tags (e.g., {\tt technology}). This
reduces the level of variability in tag usage and, ultimately, the variance of tagging.

\begin{table}[t]
    \tbl{Variance deviation from $\sigma_{max}$ for Delicious, Flickr and StumbleUpon
    \label{tab:vardev}}{
    \small
    \begin{tabular}{|c|c|c|c|}
    \hline
    $k$ & \textbf{Delicious} & \textbf{Flickr} & \textbf{StumbleUpon}\\
    \hline
    0.001        &    18.60    & 24.47         & 11.21 \\
    \hline
    0.01         &    79.82    & 61.76         & 20.87 \\
    \hline
    0.05         &    95.50    & 88.27         & 48.90\\
    \hline
    0.1         &    97.61    & 94.61         &  73.83\\
    \hline
    \end{tabular}}
\end{table}


\subsection{Variety in Tagging Behavior between Users and their Friends}
\label{sub:taggingvariety}

We study the relationships between the tagging activity of a user in a Social Sharing system
and the tagging activity of her friends.
Our ultimate goal is to determine if there exists a correlation between the number of different tags (i.e., the variety) applied by a given user and the variety of tags applied by her social contacts.
Also for this experiment we considered only Flickr and Delicious, given the absence of friending features in StumbleUpon.

We define a random variable $X$ which returns the number of friends of a given user who
have applied more tags than her. We can graphically represent this result by plotting the {\em
Cumulative Distribution Function} (CDF) $F_X(x)$ of $X$, i.e., the probability that $X$ exceeds a
fixed threshold value $x$: $F_X(x) = P(X > x)$.

\begin{figure}[ht]
	\centering
	\includegraphics[width=.49\columnwidth]{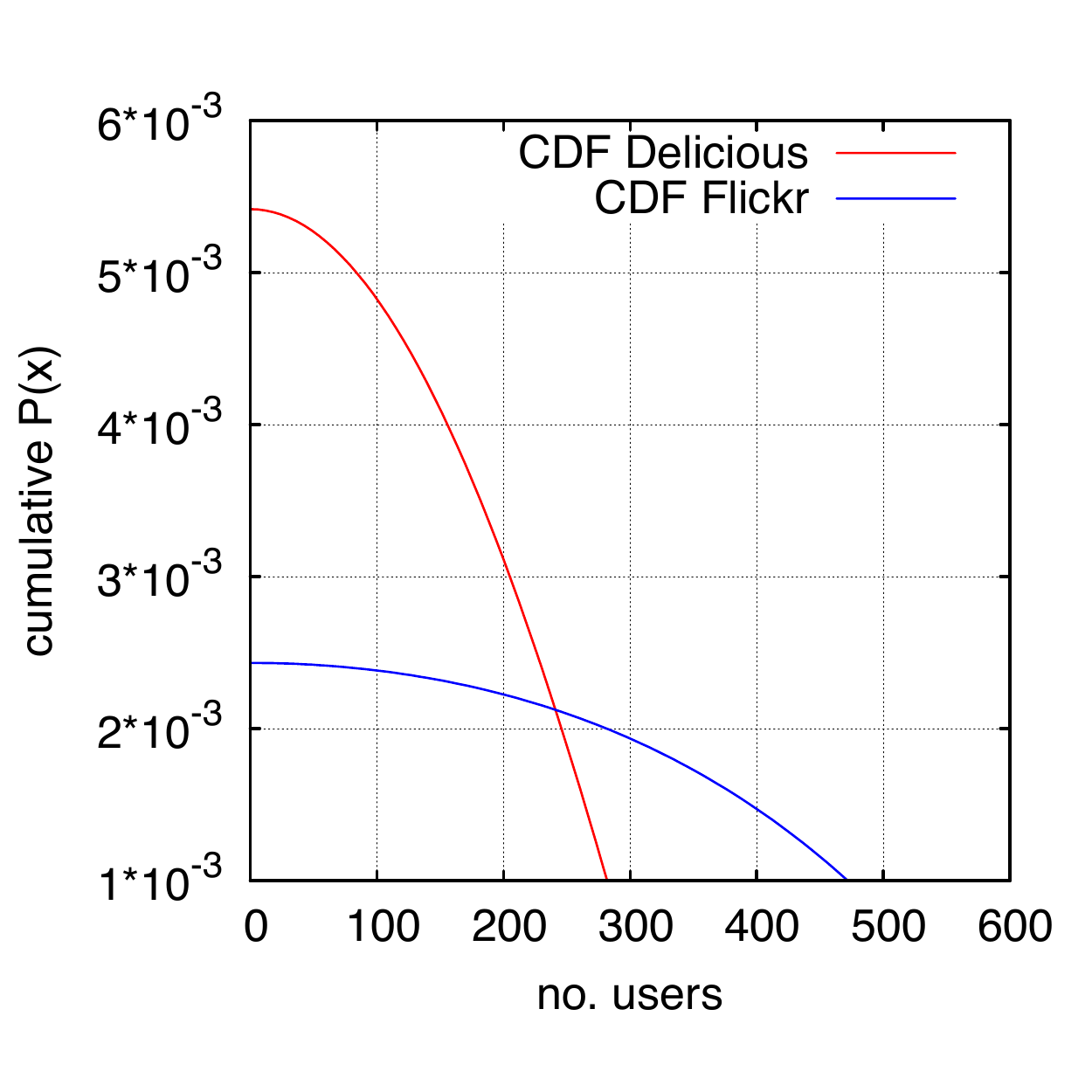}%
	\caption{Variety in Tagging: CDF in Flickr and Delicious}%
	\label{fig:cdfFD}%
\end{figure}

The CDF has been calculated as follows: first of all, for each Flickr and Delicious profile, we
calculated the overall number of tags she assigned onto the platform; then, the same operation has
been performed for all her friends. For each user $u$, it has been evaluated how many of her
friends have assigned more tags than $u$. Finally, these values have
been normalized. This allowed us to compute the CDF $F_X(u)$, which is graphically reported in
Figure~\ref{fig:cdfFD}.

From Figure~\ref{fig:cdfFD}, it emerges that the value of $F_X(x)$ is rather small both for Delicious and Flickr. Such a behavior is likely to be influenced by the heavy-tailed distribution in the frequency of tagging. In fact, since a large fraction of users tend to apply a small number of tags, we expect that, on average, the frequency of tagging of the users tends to flat around small and generally similar values. Therefore, the probability that    a user has applied more tags than her friends is generally quite small.

In addition, the CDF in Delicious approaches 0 when $x \approx 300$; by contrast, the decrease of the CDF in Flickr appears to be more soften and the it is close to $0$ when $x \approx 580$. This means that the chance that a user has applied more tags than her friends is more marked in Flickr than Delicious.

This behavior can be explained as follows: in Flickr there are several users who want to promote their contents by applying a large number of (often common) tags. For these users, the chance that their frequency of tagging is higher than that of their friends is quite high. By contrast, Delicious users apply tags because their main goal is to quickly find contents of their interest and tags make the retrieval of these contents easier. In such a case, the social aspect emerging in Flickr is less relevant than in Delicious and, therefore, there is a large number of users who apply less tags than their friends.

\subsection{Findings}
\label{sub:tagfindings}

In the following we summarize the main results reported in this section:

\begin{enumerate}

\item The entropy of the tag-based profiles depends very much on the Social Sharing platform.
For most of the users, the entropy of the Delicious profile is higher than that of StumbleUpon or Flickr ones.
The actual tag usage in Delicious is more heterogeneous than that of StumbleUpon and Flickr and therefore also harder to predict.
In addition, we do not notice any form of correlation between the entropy measures of a user's profiles across the different environments.

\item The vast majority of users tend to apply most of their tags in a uniform fashion independently of the platform in which they operate in.
However, there are some outlier tags which are much more frequently used than others.
This observation particularly holds for Flickr where users tend to -- albeit seldomly -- deviate from their usual tagging behavior.

\item The variety of the tagging activities of a user is influenced by her friending activities.
This is particularly relevant in Flickr, in which users are interested to enhance the visibility of the contents they produce.

\end{enumerate}

The previous findings can be explained as follows: in narrow folksonomies like Flickr tagging activities play a secondary role because users prefer spend their time in cultivating social relationships and tags are often perceived as a mean to promote the contents they generate. The main consequence is that user contributed tags cover some specific interests and, therefore, users tend to frequently apply a limited number of tags. This explains why, in general, tag-based profiles in Flickr feature lower values of entropy than Delicious ones.

An opposite behavior emerges in broad folksonomies: most of the user perceive tagging services as an actually relevant service and use tags to better organize the resources at their disposal. If a user diversifies her interest by posting/consumming resources of different types (e.g., she is interested both in Web pages about Computer Technology and Contemporary Art), then the tags she will contribute will reflect this diversification. This explains why tag-based profiled of Delicious users have a higher level of entropy than Flickr ones.

\section{User Profile Analysis at the semantic level}
\label{sec:userprofile-semantics}

In this section we analyze the semantics of tag-based profiles across different systems.
To this purpose, we relate tags to ontological concepts such as DBpedia or WordNet entities and transform tag-based profiles into semantic profiles.

Such a mapping is useful to derive further insights into the topics composing a given user profile.

%

\subsection{Meaningful Concepts in Tag-based User Profiles} \label{sec:userprofile-semantics-concepts}
To analyze the semantics of user profiles, we relate tags to {\em (i)} concepts of a lexical database ({\em WordNet}), and {\em (ii)} concepts of an encyclopedia ({\em DBpedia}).
WordNet allows us to characterize the tagging behavior from a lexicographic point of view because it describes what language people use to annotate resources. DBpedia  is useful to describe the ontological concepts (e.g., persons, events, and other \emph{things}) people refer to in their tagging activities.

\subsubsection{WordNet Concepts} \label{subsub:wordnet}

\begin{figure}[t!]
\begin{minipage}[tb]{\textwidth}
\begin{center}
\small
\textit{Fraction of tags that can be mapped onto WordNet categories:}
\end{center}
\end{minipage}
\subfigure[Flickr and Delicious]{
 \label{fig:wordnet-fraction-of-wordnet-tags-fd}
 \begin{minipage}[tb]{0.32\textwidth}
  \centering
  \includegraphics[width=\textwidth]{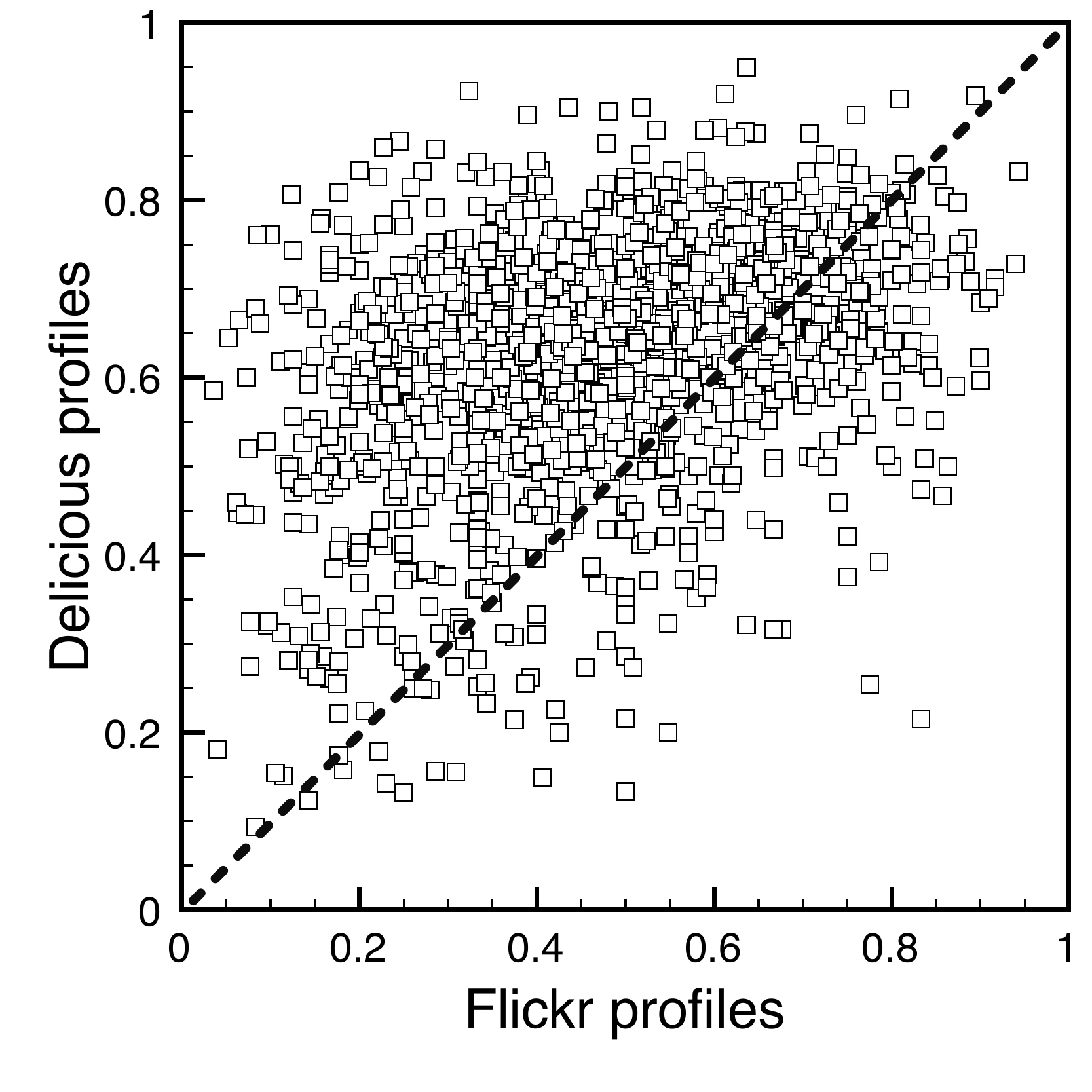}
   \end{minipage}}
\subfigure[Flickr and StumbleUpon]{
 \label{fig:wordnet-fraction-of-wordnet-tags-fs}
 \begin{minipage}[tb]{0.32\textwidth}
  \centering
  \includegraphics[width=\textwidth]{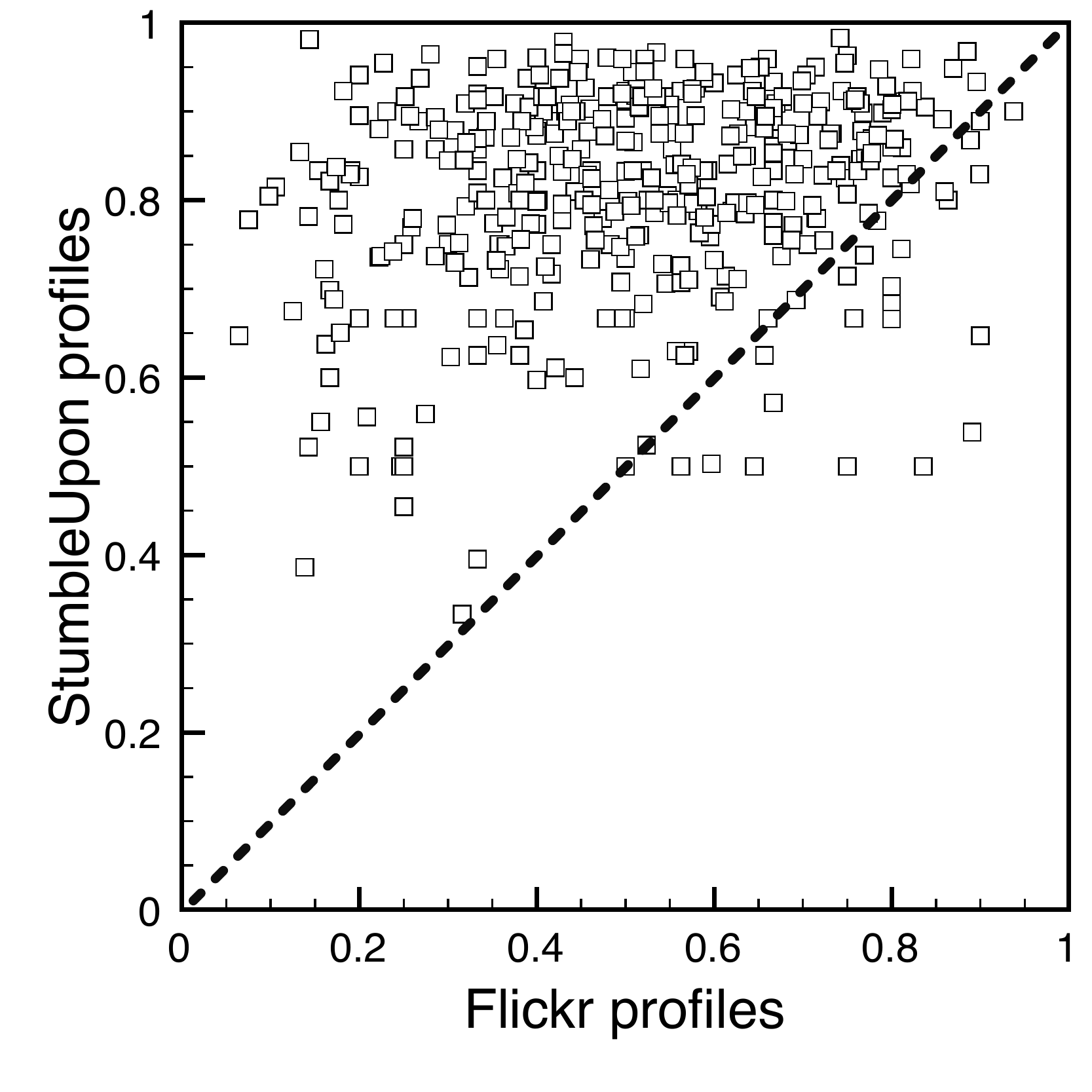}
   \end{minipage}}
\subfigure[StumbleUpon and Delicious]{
 \label{fig:wordnet-fraction-of-wordnet-tags-sd}
 \begin{minipage}[tb]{0.32\textwidth}
  \centering
  \includegraphics[width=\textwidth]{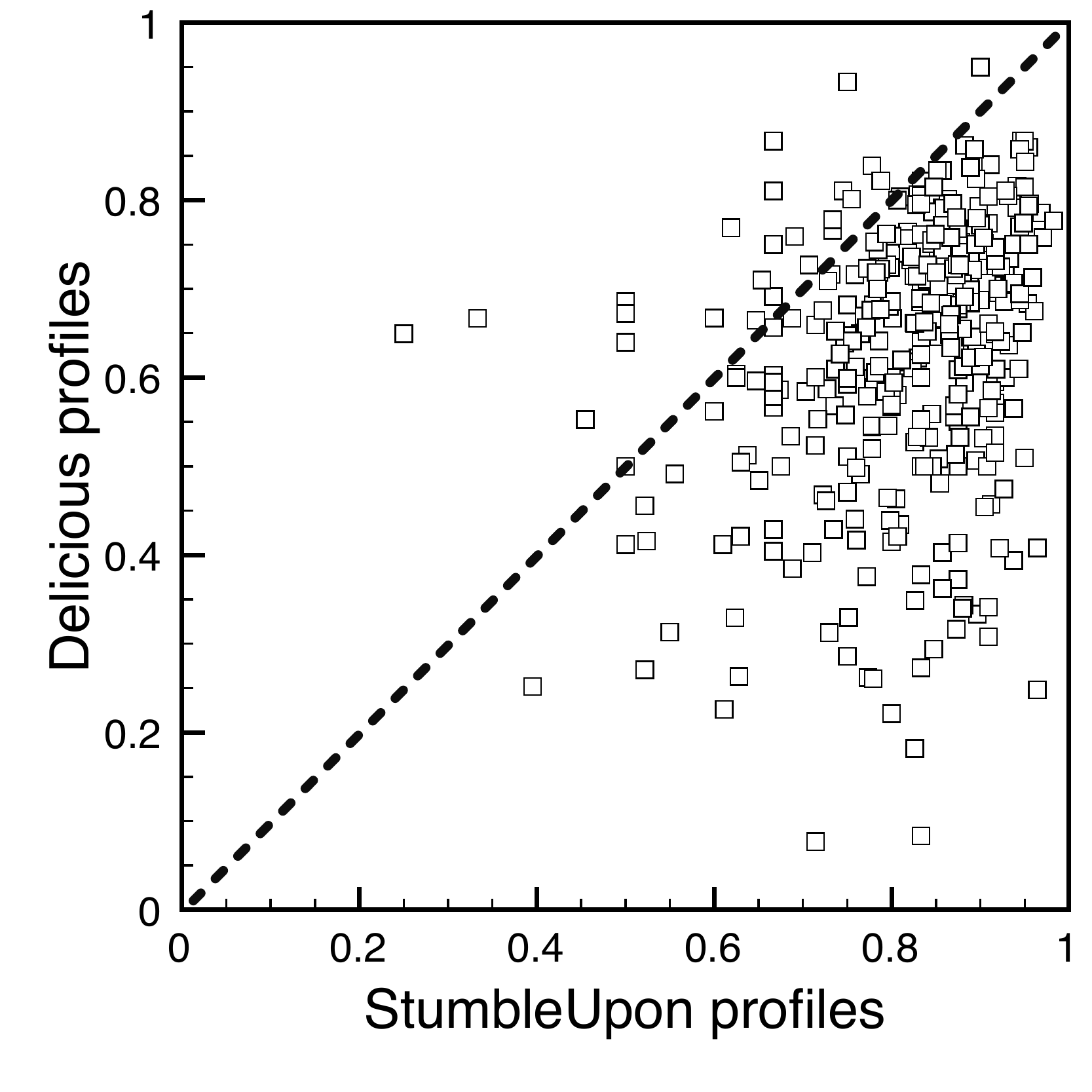}
   \end{minipage}}
\begin{minipage}[tb]{\textwidth}
 \begin{center}
 \small
\textit{Fraction of tags that can be mapped onto DBpedia entities:}
\end{center}
\end{minipage}
   \subfigure[Flickr and Delicious]{
 \label{fig:dbpedia-fraction-of-dbpedia-tags-fd}
 \begin{minipage}[tb]{0.32\textwidth}
  \centering
  \includegraphics[width=\textwidth]{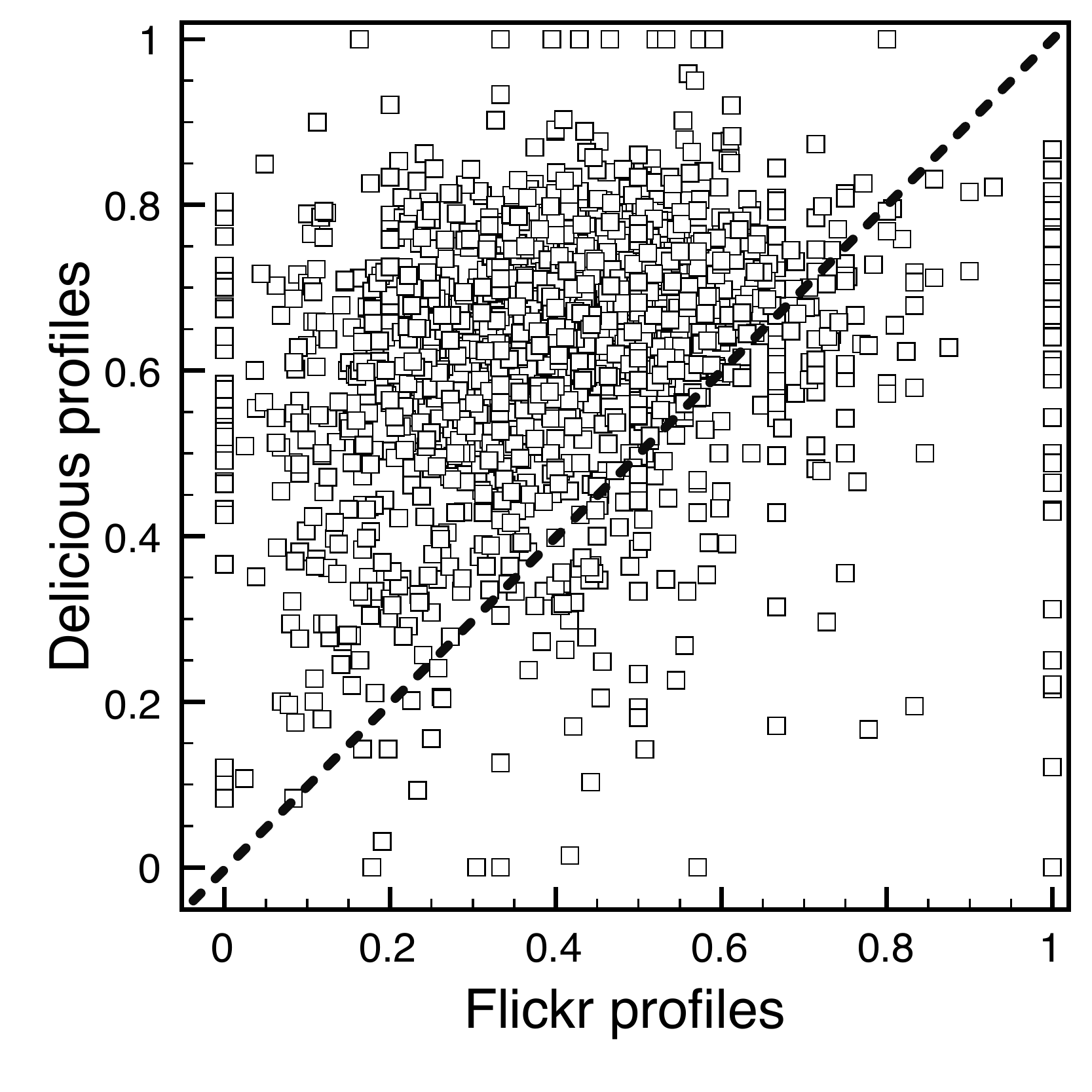}
   \end{minipage}}
\subfigure[Flickr and StumbleUpon]{
 \label{fig:dbpedia-fraction-of-dbpedia-tags-fs}
 \begin{minipage}[tb]{0.32\textwidth}
  \centering
  \includegraphics[width=\textwidth]{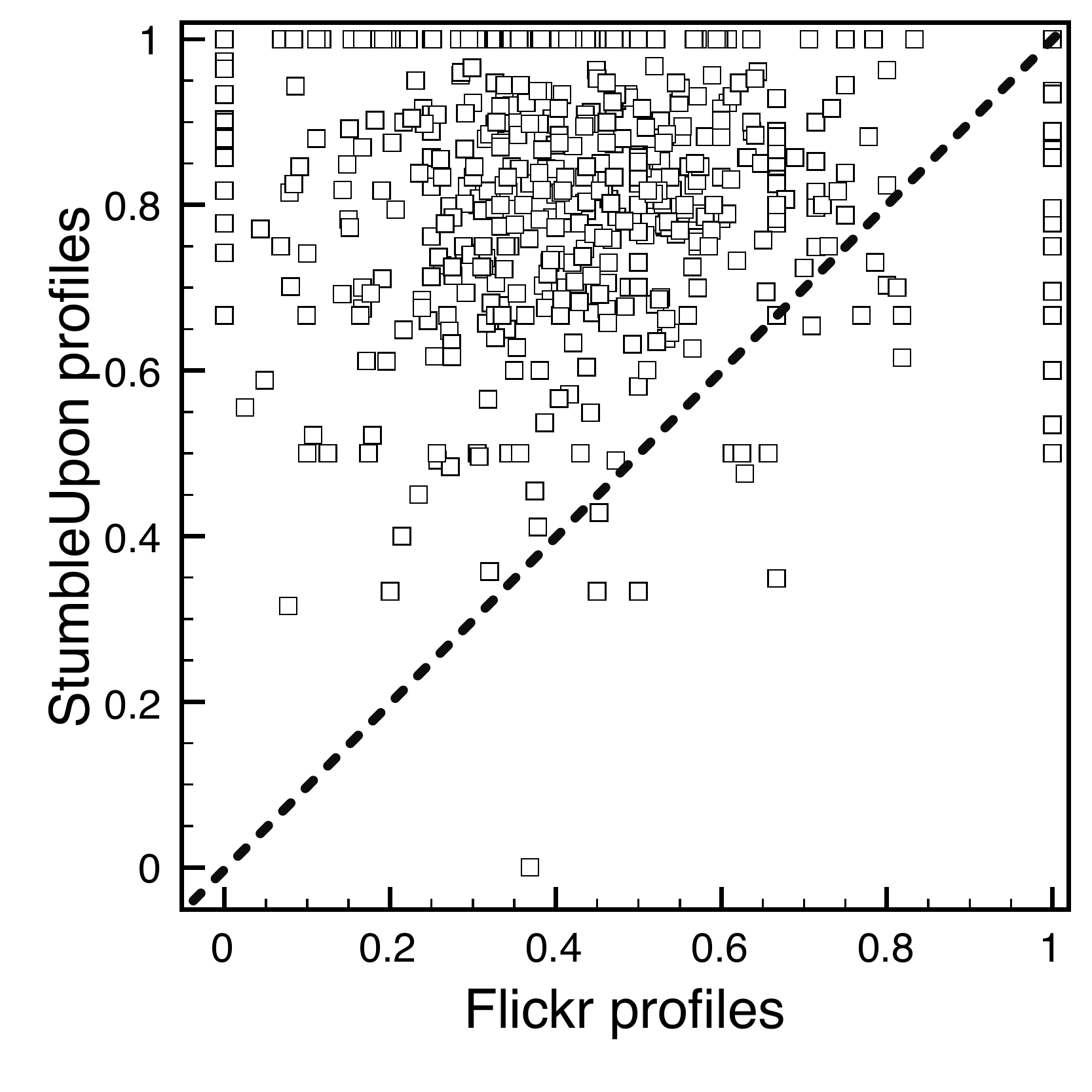}
   \end{minipage}}
\subfigure[StumbleUpon and Delicious]{
 \label{fig:dbpedia-fraction-of-dbpedia-tags-sd}
 \begin{minipage}[tb]{0.32\textwidth}
  \centering
  \includegraphics[width=\textwidth]{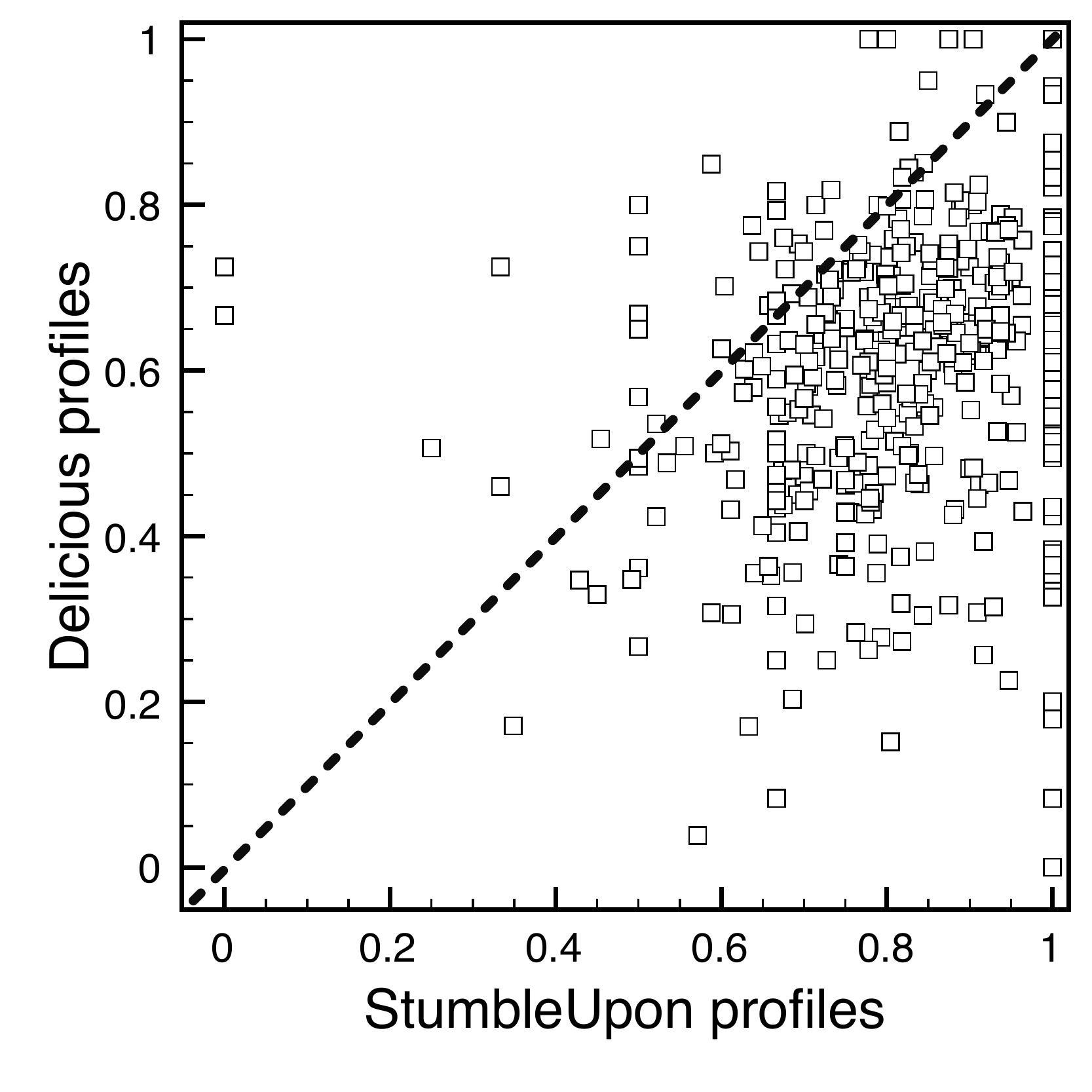}
   \end{minipage}}
   \vspace{-0.2cm}
\caption{Fraction of tags per profile that can be mapped onto (a-c) WordNet categories and (d-f) DBpedia entities. }
\label{fig:fraction-of-meaningful-tags}
\end{figure}

WordNet~\cite{Fellbaum1998} is a lexical database that groups together words that have a similar meaning into categories, so-called {\em synsets}. WordNet concepts are grouped into 45~top level categories such as nouns describing plants, food or communication, or verbs describing perceptions or motions.
Our procedure requires to map each tag in a user profile onto one of the 45~top level
concepts of the WordNet synset hierarchy (cf. lexicographer files\footnote{\url{http://wordnet.princeton.edu/man/lexnames.5WN.html}}).
Such a mapping can be successful (if we find a WordNet category corresponding to that tag) or not.

The {\em degree of alignment} of a tag-based user profile can be defined as the percentage of tags in the profile that can be mapped onto WordNet categories.
We considered Social Sharing systems in a pairwise fashion and, for each user registered to both the two systems we computed the percentage of tags which have been successfully mapped onto WordNet categories.
This procedure generated diagrams reported in Figures~\ref{fig:fraction-of-meaningful-tags}(a-c).
So, for instance, in Figure~\ref{fig:fraction-of-meaningful-tags}(a) each dot represents a user $u$.
The abscissa (resp., ordinate) of this dot is the percentage of tags in the Flickr (resp., Delicious) profile of $u$ successfully mapped onto WordNet categories.

Figures~\ref{fig:fraction-of-meaningful-tags}(a-c) allow us to infer (1) to what extent a user
applies proper words as tags that one can also find in dictionaries, and (2) whether this usage
pattern is rather user-specific or depends (also) on the social tagging environment.

Figure~\ref{fig:wordnet-fraction-of-wordnet-tags-fd} characterizes Flickr and Delicious user profiles and shows that the success of mapping tags to WordNet concepts  varies strongly among the different user profiles.
If the success would merely depend on the user then one would expect that the large majority of the dots would be plotted close to the diagonal line.
In such a case, in fact, the percentage of successfully mapped tags in Delicious would be equal to the percentage of successfully mapped tags in Flickr.
However, the figure reveals that, on average, 62.13\% of the Delicious profile tags can be related to WordNet categories in comparison to 47.76\% for Flickr profiles.

For StumbleUpon profiles, the percentage of successfully mapped tags is 81.3\% or higher (see Figures~\ref{fig:wordnet-fraction-of-wordnet-tags-fs}
and~\ref{fig:wordnet-fraction-of-wordnet-tags-sd}). Both figures indicate that the tag vocabulary people utilize in
StumbleUpon seems to be less noisy than their vocabulary in Flickr or Delicious. This observation
can be explained by the fact that people who annotate
Web resources in StumbleUpon receive category-based tagging suggestions. For example, they can
select from categories such as \emph{entertainment} or \emph{technology}, hence terms which are
also contained in thesauri such as WordNet.

There are only a few users for whom the Flickr or Delicious profiles seem to feature less noise than their StumbleUpon profiles.
The definition of \emph{noise} depends on the application context.
Here, we consider those tags that do not appear in thesauri such as WordNet as noise.
Indeed, those tags might still be of interest to certain applications that consume tag-based user profiles.
Therefore, in the following we investigate tags related to ontological concepts (DBpedia resources).

\subsubsection{DBpedia Concepts}
\label{subsub:dbpedia}

DBpedia~\cite{dbpedia/iswc07} provides RDF representations for the majority of concepts that are
explained in the Wikipedia encyclopedia. DBpedia concepts are classified according to the DBpedia
ontology\footnote{\url{http://dbpedia.org/ontology/}} which features more than 50~classes such as
\emph{Person}, \emph{Place} or~\emph{Work}. To relate tag-based profiles to DBPedia concepts, we
exploit three named entity recognition services: DBpedia Spotlight\footnote{\url{http://dbpedia.org/spotlight}},
Alchemy\footnote{\url{http://alchemyapi.com}} and OpenCalais\footnote{\url{http://opencalais.com}}.
These services detect entities in a given text snippet and link the identified entities to
ontological concepts (e.g., DBPedia concepts) if possible. In our analysis, we serialize a
tag-based profile to a document and then process the document via the different services. We
aggregate the entities extracted from these services to increase the recall of detected entities.
The accuracy of the named entity extraction varies slightly among the services. For example, tested
on a news corpora from \emph{New York Times}, Mendes et al. report about precision for Alchemy and
DBpedia spotlight that ranges between 70\% and 80\%~\cite{isem2011mendesetal}.

In Figures~\ref{fig:fraction-of-meaningful-tags}(d-f) we plot the fraction of tags that can be
mapped onto DBpedia resources. The meaning of these diagrams is as follows: we considered Social
Sharing systems in a pairwise fashion and each dot identifies a user. The abscissa (resp.,
ordinate) of a dot is the percentage of tags in the user profile associated with the first
(resp., the second) Social Sharing system successfully mapped onto a DBPedia concept.

Results emerging from Figures~\ref{fig:fraction-of-meaningful-tags}(d-f) confirm our results in the context of the WordNet analysis.
Delicious and StumbleUpon profiles allow for better alignment with ontological concepts than tag-based Flickr profiles.
On average, we observe that 41.1\% of the tags in a Flickr profile can be matched with a DBpedia entity.
For Delicious and StumbleUpon profiles, this fraction raises to 61.5\% and 80.9\%, respectively.

However, as indicated in Figure~\ref{fig:dbpedia-fraction-of-dbpedia-tags-fd}, there are some
users (3\% of all Flickr users) for whom all elements in the tag-based Flickr profiles can be mapped onto DBpedia concepts.
For these users, we observe that their tagging vocabulary is very limited because it consists, on average, of only 2.55~distinct tags.
Figure~\ref{fig:dbpedia-fraction-of-dbpedia-tags-fs} reveals that there is a similar behavior on StumbleUpon: for 15.6\% of the StumbeUpon profiles we find that 100\% of the tags in these profiles have a match on DBpedia. The average size of the corresponding profiles is again
small and contains, on average, only 8.43~distinct tags.

Moreover, Figures~\ref{fig:fraction-of-meaningful-tags}(d-f) indicate that the tagging
behavior of a user varies across the different tagging environments. For example, if a user in Delicious, applies
tags, which can be successfully aligned with semantic concepts from DBpedia, then this does not necessarily imply a similar behavior in Flickr.

\subsection{Variety of Semantic User Profiles}\label{sec:userprofile-semantics-concept-variety}

In order to study the thematic variety of user profiles, we analyze the usage characteristics with
respect to the different \emph{types of concepts} that relate to users tagging activities. We
utilize again variance (cf. Section~\ref{sub:tagvarianceusage}) as a measure to quantify whether a
user often refers to the same concept types and topics (low variance) or, vice versa, whether her
tagging activities relate to multiple topics (high variance). The sample variance
$\sigma_{C, u}$ describing the variety of topics $C$ to which a user $u$ refers to is

\begin{equation}\label{eq:concept-variance}
\sigma_{C, u} = \frac{\sum_{c_j \in C_u} \left( f_u(c_j) - \mu_{C, u} \right)^2 }{|C_u| -1}
\end{equation}

Here, $C_u$ is the set of distinct concept types that relate to at least one of the tags provided
by $u$. The frequency $f_u(c_j)$ denotes the number of tags contributed by $u$ that are related to
a topic $c_j$ 
and $\mu_{C, u}$ is the average number of tags per topic.

To detect topics we used both WordNet and DBPedia.

\subsubsection{WordNet Concept Types}
\begin{figure}[t!]
\begin{minipage}[tb]{\textwidth}
 \begin{center}
 \small
\textit{Variance of references to certain types of WordNet categories:}
\end{center}
\end{minipage}
\subfigure[Flickr and Delicious]{
 \label{fig:wordnet-variance-wordnet-type-usage-fd}
 \begin{minipage}[tb]{0.32\textwidth}
  \centering
  \includegraphics[width=\textwidth]{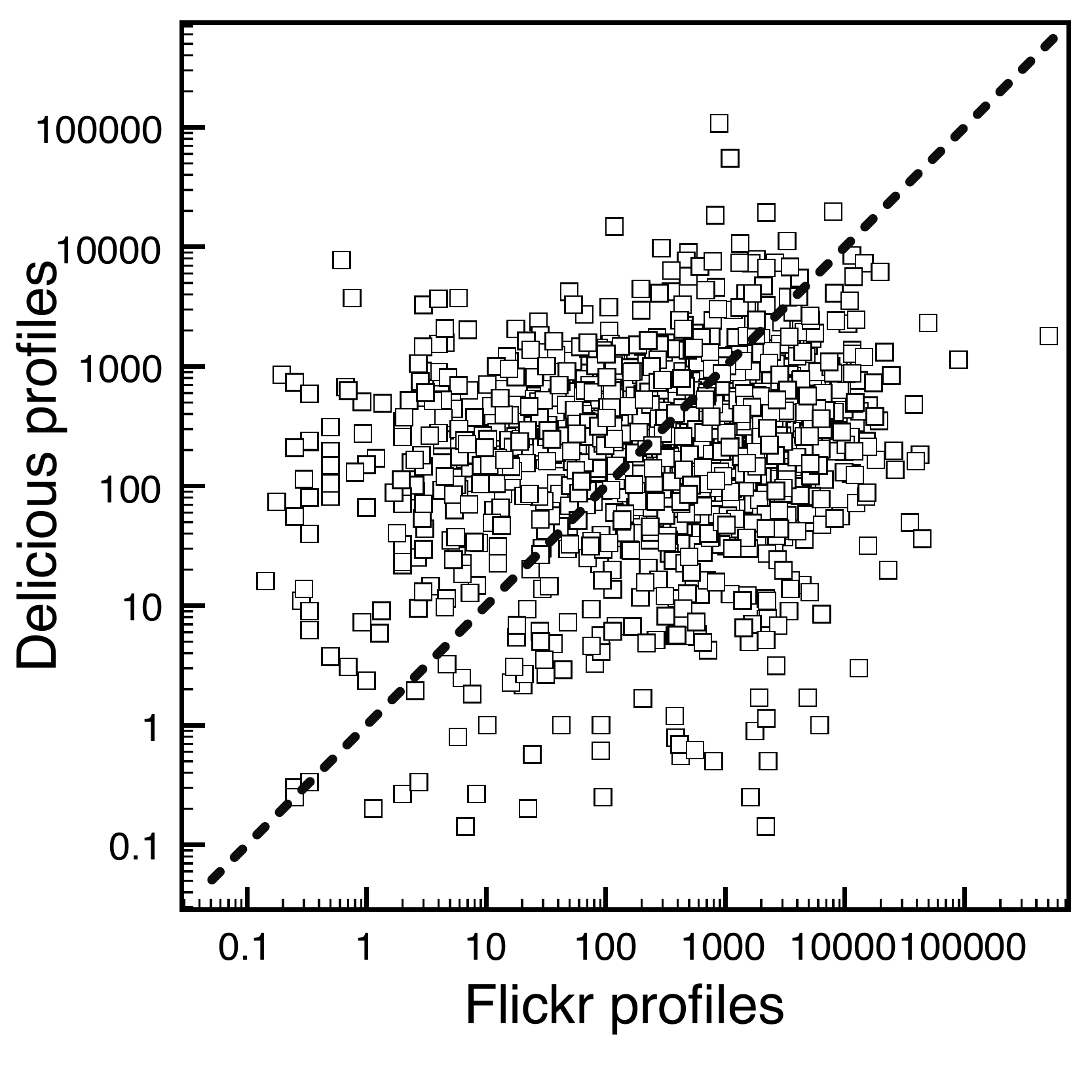}
   \end{minipage}}
\subfigure[Flickr and StumbleUpon]{
 \label{fig:wordnet-variance-wordnet-type-usage-fs}
 \begin{minipage}[tb]{0.32\textwidth}
  \centering
  \includegraphics[width=\textwidth]{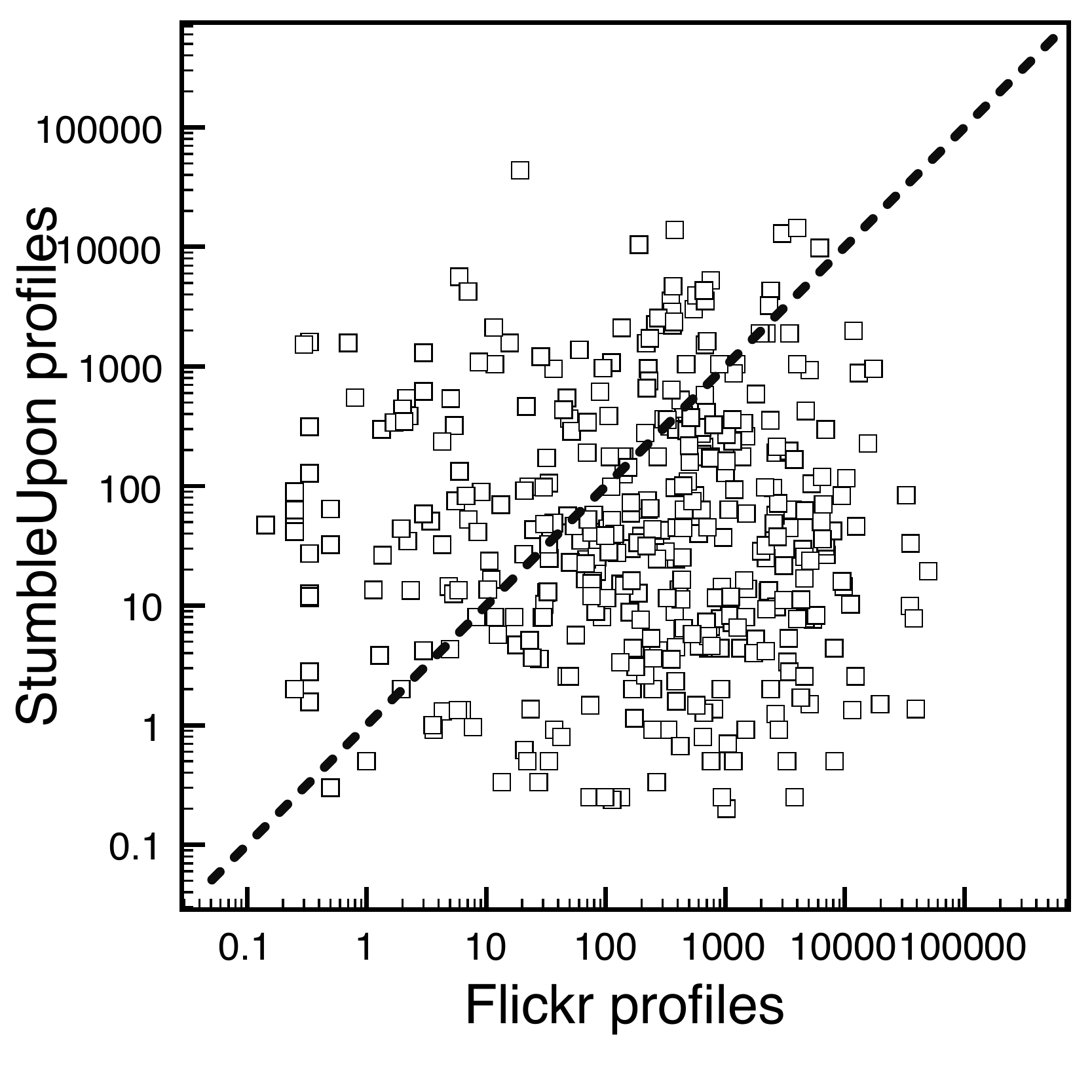}
   \end{minipage}}
\subfigure[StumbleUpon and Delicious]{
 \label{fig:wordnet-variance-wordnet-type-usage-sd}
 \begin{minipage}[tb]{0.32\textwidth}
  \centering
  \includegraphics[width=\textwidth]{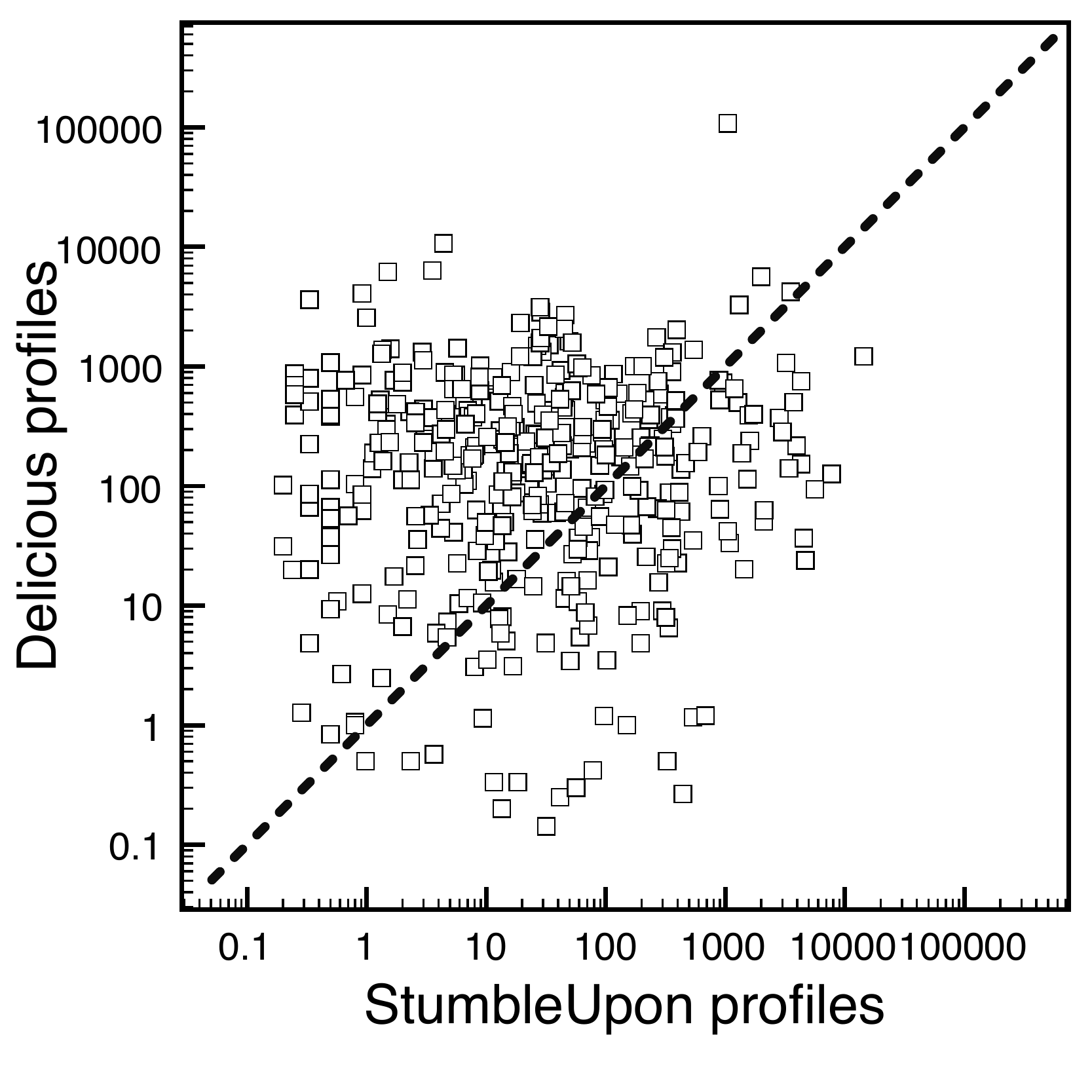}
   \end{minipage}}
\begin{minipage}[tb]{\textwidth}
 \begin{center}
 \small
\textit{Variance of references to certain types of DBpedia concepts:}
\end{center}
\end{minipage}
   \subfigure[Flickr and Delicious]{
 \label{fig:dbpedia-variance-dbpedia-type-usage-fd}
 \begin{minipage}[tb]{0.32\textwidth}
  \centering
  \includegraphics[width=\textwidth]{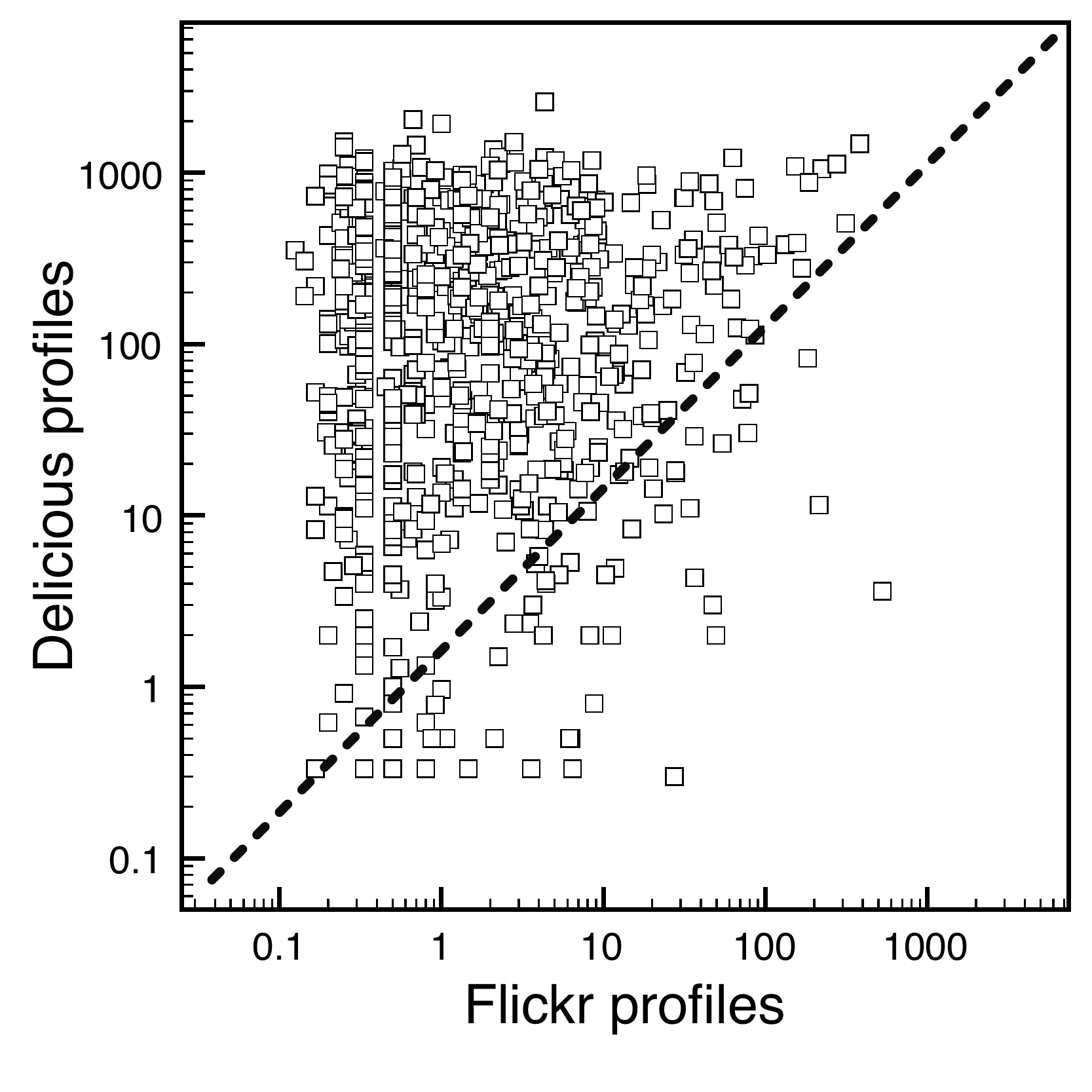}
   \end{minipage}}
\subfigure[Flickr and StumbleUpon]{
 \label{fig:dbpedia-variance-dbpedia-type-usage-fs}
 \begin{minipage}[tb]{0.32\textwidth}
  \centering
  \includegraphics[width=\textwidth]{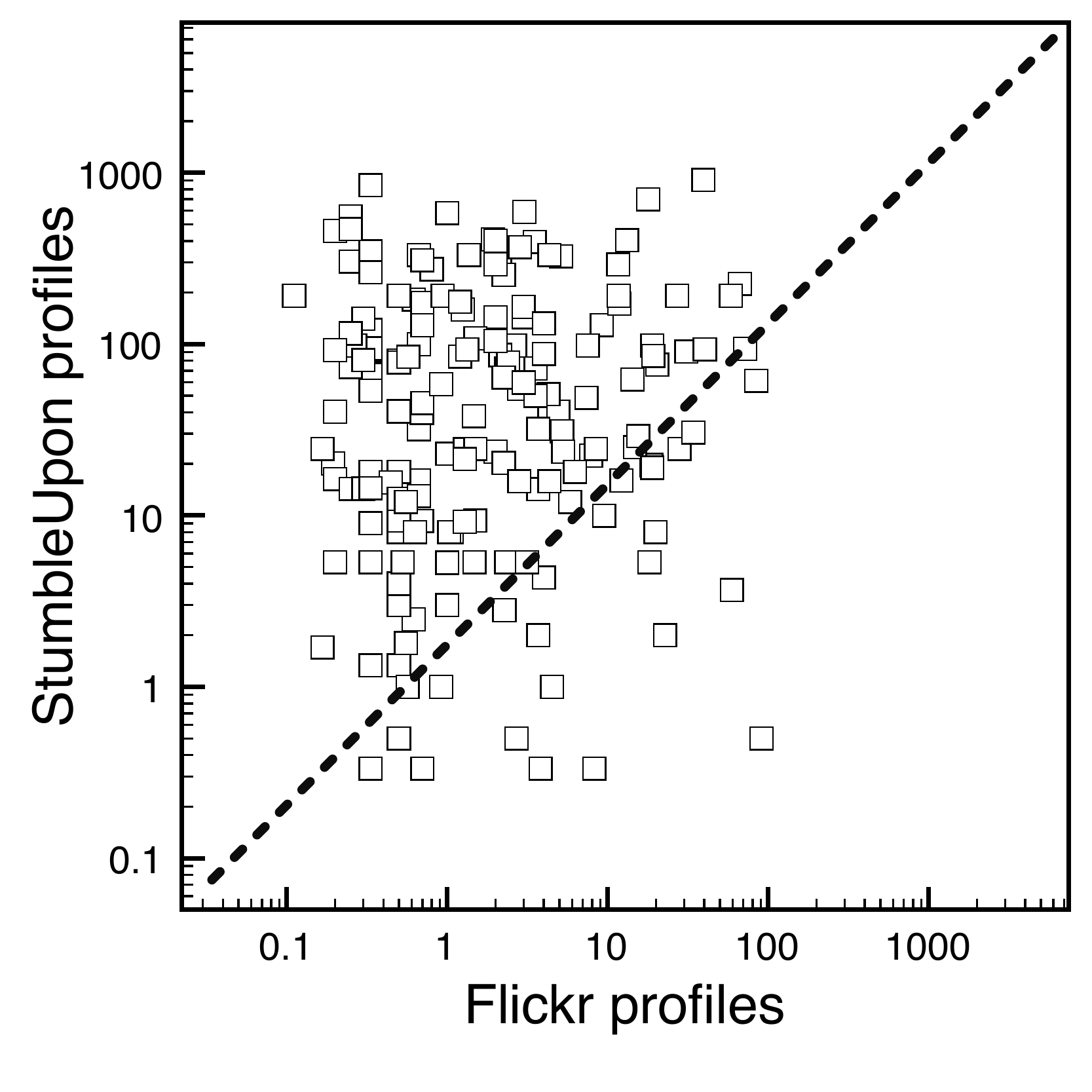}
   \end{minipage}}
\subfigure[StumbleUpon and Delicious]{
 \label{fig:dbpedia-variance-dbpedia-type-usage-sd}
 \begin{minipage}[tb]{0.32\textwidth}
  \centering
  \includegraphics[width=\textwidth]{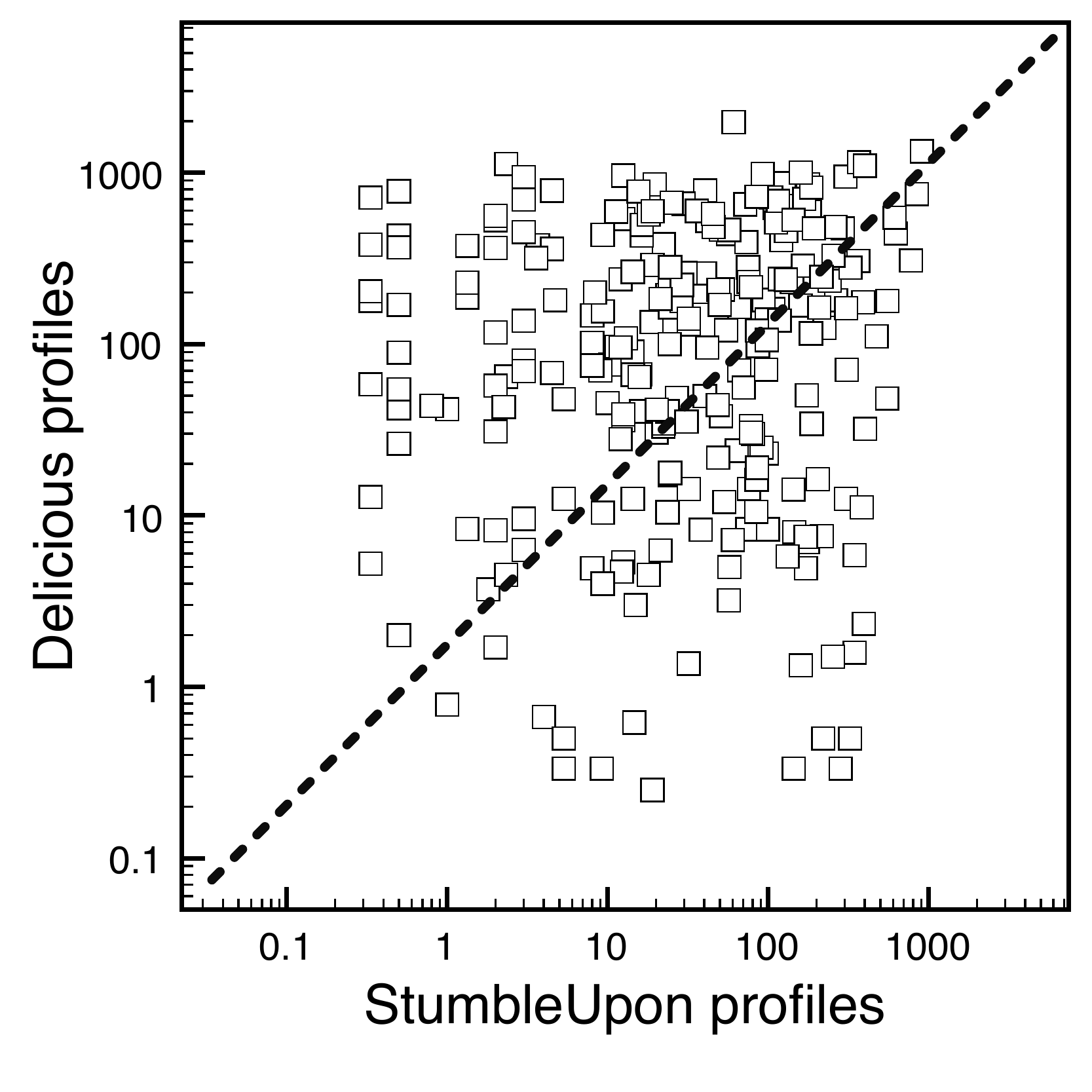}
   \end{minipage}}
 \vspace{-0.2cm}
\caption{Variance of references to certain types of (a-c) WordNet categories and (d-f) DBpedia concepts.}
\label{fig:semantics-variance-type-usage}
\end{figure}

In Figures~\ref{fig:semantics-variance-type-usage}(a-c) we compared the variance in the topic usage by mapping user tags onto WordNet categories.

Figures~\ref{fig:semantics-variance-type-usage}(a-c) indicate that there is no significant
difference between the topic varieties of user profiles in Flickr and Delicious. In fact, for
53.3\% of the users, the topic variety of their Flickr profiles is higher than the ones of their
Delicious profiles while for 46.7\% it is the vice versa.

It is interesting to observe that the individual behavior differs across the systems. For example,
some users are concerned with various types of topics in Flickr but only with a few types of topics
in Delicious or vice versa.

Between Flickr and Delicious the differences in the design of the tagging environments do not seem
to impact the variety of the types of concepts people refer to in their tagging activities.
However, as depicted in Figure~\ref{fig:wordnet-variance-wordnet-type-usage-fs} and
Figure~\ref{fig:wordnet-variance-wordnet-type-usage-sd}, the StumbleUpon environment seems to
influence the tagging behavior from a thematic point of view, stronger than the individual usage
preferences of a user. About 64\% of the people feature more diversity in their Flickr profiles in
comparison to their StumbleUpon profile. When comparing StumbleUpon and Delicious profiles, we
observe that even more than 70\% of the users reveal more diverse topics in the Delicious profile that was
not generated in StumbleUpon.

\subsubsection{DBpedia Concept Types}

In Figures~\ref{fig:semantics-variance-type-usage}(d-f) we studied the variance in the topic usage by mapping user tags onto DBpedia concepts. These results clearly differ from previous results based on WordNet categories.

Overall, we observe that the variance is much smaller compared to variance measured for the WordNet
categories. On average, the variance with respect to the types of DBpedia concepts referenced by
the users is 5.9 (Flickr), 249.6 (Delicious) and 78.9 (StumbleUpon). By contrast,
the analysis of the usage variety of WordNet categories revealed, on average, a variance of 2080.9
(Flickr), 721.9 (Delicious) and 287.6 (StumbleUpon) for these services. We observe that
regarding the usage of ontological concepts (DBpedia), the tagging behavior is -- independently of
the Social Sharing system -- more uniform than the lexicographic dimension (WordNet).
Hence, the language that people use to describe resources varies more strongly than the types of
concepts to which they refer to.

Furthermore, the comparison of the tagging behavior across the different tagging environments leads
to new insights. For example, while the variety of the language differs only slightly between
Flickr and Delicious (see Figure~\ref{fig:wordnet-variance-wordnet-type-usage-fd}), we discover
that the variety of ontological concepts that people refer to strongly differs across these two
platforms as well as between Flickr and StumbleUpon. On Flickr, the variance of concept types
is -- for the majority of the users -- much smaller than on Delicious or StumbleUpon (see
Figure~\ref{fig:dbpedia-variance-dbpedia-type-usage-fd} and
Figure~\ref{fig:dbpedia-variance-dbpedia-type-usage-fs}), which indicates that, from an ontological
point of view, the tagging behavior is more uniform on Flickr than on Delicious or StumbleUpon. The
ontological analysis refines the insights from the lexicographic analysis: there exists a large
variety of lexicographic concepts to describe pictures on Flickr but the variety of the ontological
concepts that people refer to is smaller in Flickr and much broader on Delicious and StumbleUpon.

\subsection{Findings}
\label{sec:userprofile-semantics-findings} The semantic analysis revealed how the tagging
behavior of users differs across Social Sharing platforms with respect to the lexicographic and
ontological dimension. In summary, we can answer the research questions raised:

\begin{enumerate}
  \item Semantic concepts can be successfully inferred from the tag-based profiles of the
      different Social Sharing platforms. However, the success in deducing the semantics mainly
      depends on the social tagging platform rather than on the individual user behavior. For example,
      the fraction of tags that can be mapped to DBpedia concepts is 61.5\% and 80.9\% for
      Delicious and StumbleUpon respectively, while for Flickr profiles this fraction is just
      41.1\%.
  \item The semantic variety of user profiles clearly differs across Social Sharing platforms.
      Here, the results of the lexicographic and ontological analysis complement each other and
      imply the following conclusions. The variety of lexicographic concepts that people use
      within the different social tagging environments does not vary strongly between
      environments such as Flickr and Delicious. Tagging support as offered by StumbleUpon by
      means of category-like tag suggestions results in a lower lexicographic variety. However,
      the variety of ontological concepts that people refer to strongly depends on the system we consider: on Flickr we observe a more uniform behavior than on Delicious
      and StumbleUpon.
\end{enumerate}



A possible explanation is that users can apply different tags which are related to the same topic.
The number of topics is much less than the number of tags and, then, we expect that the deviation on the number of topics a user
focuses on and the average number of topics per user is small. This explains the low variance in topic distribution.
Of course, two different users can focus on the same topics but they can use different tags referring to the same topic. Due to the homophily principle, it is likely that users sharing some interests can get in touch/become friends and expand their vocabularies by absorbing the tags applied by other users. Therefore, we are not surprised at the existence of users who applied a number of tags which is much higher or much lower than the average number of tags per user. This behavior has been already highlighted in Figure \ref{fig:tas} in which a right-skewed distribution emerged in tag frequency assignment and this explains why, at the tag level, we are in presence of a high variety.

\section{Applications} \label{sec:applications}

In this section we discuss how the findings of this work can be used in practical scenarios.

The analysis carried out in this paper showed that the design features and the goals of a Social
Sharing system can influence the profile of a user. In the following we shall consider only the tagging behavior of a user because user-contributed tags can be used to infer user preferences. Due to these reasons, given a user $u$ affiliated to multiple
Social Sharing systems $\mathcal{S}_1, \ldots, \mathcal{S}_k$ we consider her tag-based profiles $P_T^1, P_T^2, \ldots, P_T^k$ in each system. We can expect that
there are some user preferences/features which are recurrent across all profiles, whereas there is
other information which is present only in one or few profiles. Therefore, each profile $P_T^i$ can
be decomposed into two components called {\em persistent} and {\em contingent} components.

The persistent component reflects user preferences which are present in multiple profiles of $u$,
and, therefore, are independent of the specific platform in which a user is registered to. The
contingent component denotes preferences of $u$ which are distinctive of a specific platform and
that are not present in other platforms; therefore, these preferences depend on the specific goals
and design of the platform itself.

The decomposition into persistent/contingent components has many practical applications which are detailed below.

{\em Merging of User Profiles}. Some authors studied how to merge the tags contributed by users in different platforms to build a {\em global user profile} \cite{Szomszor08}. Users may apply different tags to describe the same concept ({\em synomymous tags}) and, therefore, tag alignment techniques to detect and merge synonymous tags are required. In these approaches, the fact that a tag may be occasionally applied in a platform or, by contrast, it identifies a user interest stated in multiple platform, is neglected. Therefore, we can think of more refined procedures to merge user profiles in which the persistent and contingent components are identified; after that, we can think of {\em merging} only the contingent components.

{\em Dealing with the cold start problem}. A relevant application of our findings is in the area of Recommender Systems and it is related to the ability of effectively dealing with the {\em cold start problem}. Such a problem arises each time a user $u$ subscribes for the first time to a system and her profile is empty. In such a case it is not possible to provide $u$ with personalized recommendations. Tags in the persistent component of the profile of $u$ represent well-defined user preferences and they can be used to recommend contents to $u$. Such a knowledge is particularly useful in the context of {\em content-based} recommender systems \cite{de2008integrating} because we can
use tags in the persistent component as a form of {\em bootstrap knowledge}. In particular, we envision the possibility of comparing the tags describing a resource with the tags in the persistent profile of $u$ to determine whether such a resource may be of interest to $u$.

{\em Computing User Similarities}. Since the inception of Collaborative Filtering techniques, many authors studied the
problem of computing the similarity degree of two users. Such a problem has been recently considered in the context of folksonomies \cite{DeQuUr10}. In these approaches the computation of the similarity degree of two users is translated to the computation of the similarity of the
tags appearing in their profiles.

As showed in Section \ref{sub:general-definitions}, tags can be {\em weighted} and the weight of a tag reflects its relevance in defining user interests. Tag weights could be used to compute in a more precise fashions pairs (or groups) of similar users \cite{DeQuUr10}.


If a user $u$ enters for the first time in a new Social Sharing system, her tag-based profile is empty and we can not identify her neighborhoods and use the information they contributed to recommend resources to $u$. We propose to use tags in the persistent component of the profile of $u$ along with their weights to identify groups of similar users: in particular, given a Social Sharing system $\mathcal{S}_i$, and a user $v$ in $\mathcal{S}_i$, we suggest to compare the tags (and their weights) in the profile of $v$ with the tags (and associated weights) in the persistent component of the profile of $u$ to determine the similarity degree of $u$ and $v$.

To compute the weight of a tag $t$ we assume that a tag-ranking algorithm $\mathcal{R}$ is available and that $\mathcal{R}$ associates each tag $t$ with a weight $w_t$. After this we consider two cases: in the former case $u$ has applied $t$ in {\em exactly one} of the available Social Sharing systems, say $\mathcal{S}_i$. In this case $t$ is part of the contingent component of the profile of $u$ and its weight is equal to $w_i(t) = \mathcal{R}(t)$; here the subscript $i$ informs us that $t$ appeared in $\mathcal{S}_i$.

In the latter case, we suppose that $u$ has applied $t$ in more than one Social Sharing systems (or in more than a certain number). In such a case $t$ can be regarded as part of the persistent component of the profile of $u$. More formally, a threshold $\tau$ can be experimentally defined and we say that $t$ is part of the persistent component of the profile of $u$ if it has been exploited by $u$ in at least $\tau$ Social Sharing systems.
In this case we associate $t$ with an array $\mathbf{w} = [w_1(t), \ldots, w_n(t)]$ and each $w_i(t)$ is computed as in the former case. Of course $w_i(t) = 0$ if $t$ does not appear in $\mathcal{S}_i$. The weight associated with $t$ can be computed by aggregating $w_1(t), \ldots, w_n(t)$
into a global value. There are several strategies to aggregate weights and some of them have been extensively studied in the field of {\em contextual recommender systems}: for instance, we can use operators like {\tt avg} and {\tt sum} \cite{Adomavicius*05}.

\section{Related Works}\label{sec:related}
Research related to \emph{social network analysis} and social tagging acquired a lot of attention during
the last decade. In this paper, we focused on the analysis of individual user profiles that can
be inferred from social networking and tagging activities and studied the characteristics of these
profiles across system boundaries. Our findings are of particular interest to generic user modeling
systems,
that aim to re-use profile information in different application contexts.

In this section, we discuss related literature from three main perspectives: {\em (i)} the analysis
of user data that is distributed across Social Web systems, {\em (ii)} the problem of identifying
users on the Social Web and {\em (iii)} approaches to interpreting the semantics of tags and
categorize them.

\subsection{User Data Analysis across Social Web Systems} \label{sub:analysisuserdata}

In the latest years many researchers studied how to aggregate data coming from disparate Social Web
systems with the goal of enhancing the level of personalization offered to the end users
\cite{crossSystem/umap/2011/ShiLH11,crossSystemUM/SocialWeb/UMUAI/abel/2011} or improve the
description of Web resources \cite{ernesto/crossTagging/Hypertext2009}.

To the best of our knowledge, one of the first approaches targeting at merging data residing on
different folksonomies is the already mentioned work by~\cite{Szomszor08}.

While the approach of~\cite{Szomszor08} focuses on the construction of global user profiles, other
authors observe that the process of aggregating tagging data can be beneficial not only to produce
more detailed user profiles but also to better describe resources available in a Social Web system. A
relevant example is provided in \cite{ernesto/crossTagging/Hypertext2009}; in that paper the
authors consider users of blogs concerning music and users of {\tt Last.fm}, a popular folksonomy
whose resources are musical tracks. The ultimate goal of \cite{ernesto/crossTagging/Hypertext2009}
is to enrich each Social Web system by re-using tags already exploited in other environments.
This activity has a twofold effect: it first allows the automatic annotation of resources which
were not originally labeled and, then, enriches user profiles in such a way that user similarities
can be computed in a more precise way.

Starting from the findings of \cite{Szomszor08} and \cite{ernesto/crossTagging/Hypertext2009},
many researchers studied how to get real benefits from the aggregation of social data. A
popular research trend consists of exploiting aggregated social data to generate high quality
recommendations. For instance, in \cite{crossSystem/umap/2011/ShiLH11}, the authors suggest to
use tags present in different Social Web systems to establish links between items located in
each system.

In \cite{DBLP:conf/ideas/MeoNQRU09} the authors show how to merge ratings provided by users in
different Social Web platforms to compute reputation values (which are subsequently used to
generate recommendations).

In \cite{crossSystemUM/SocialWeb/UMUAI/abel/2011} the system Mypes is presented. Mypes supports the
linkage, aggregation, alignment and semantic enrichment of user profiles available in various
Social Web systems, such as Flickr, Delicious and Facebook.

The approaches presented in this section neglect social relationships whereas, in our approach, the social and tagging behavior of a user play an equally relevant role to infer user preferences. Due to these reasons, in this paper we did not investigate new techniques for recommending items/tags in social system (which is the core of another research line \cite{ferrara2011improving}), but aimed at finding potential form of correlation between the social and the tagging behavior of a user.

\subsection{User Identification across Social Web Systems} \label{sub:user identification}

Recently, some authors focused on the problem of matching the multiple accounts created by a user in different Social Web systems ({\em user identification}). Such a research line can be considered complimentary to ours because some authors observed that the analysis of the
tag-based user profiles in different systems is useful to reveal whether two user accounts refer to
the same user or not \cite{userIdentification/socialTagging/icwsm/2011}. This kind of analysis
resembles the empirical analysis on tag-based profiles presented in this paper.

To the best of our knowledge, the first attempt to match user identities was proposed in
\cite{francesca/UserIdentification}. In that paper, the authors performed an in-depth analysis of
25 adaptive systems with the goal of identifying what user profile attributes were frequently
recurrent in each system; the outcome of such an analysis led to isolate attributes like username,
name, location and e-mail address.

More recently, other authors studied the problem of user identification in the context of Social
Networks. One of the first studies was proposed in \cite{IJWA:2009voseck}. In that paper the
authors focused on Facebook and StudiVZ\footnote{{\tt http://studivz.net/}} and investigated which
profile attributes can be used to identify users.

In \cite{zafarani2009connecting}, the authors studied 12~different Social Web systems (like
Delicious, Flickr and YouTube) with the goal of finding a mapping with the different user accounts.
This mapping can be found by applying a traditional search engine.

In \cite{userIdentification/socialTagging/icwsm/2011}, the authors suggest to combine profile attributes (like usernames) with an analysis of the tags contributed by them to identify users.
They suggest various strategies to compare the tag-based profiles of two users.
Experiments provided in \cite{userIdentification/socialTagging/icwsm/2011} are able to achieve an accuracy of almost 80\% in user identification.

Our work introduces some novelties against the approaches presented in this section. First of all,
we suggest that social relationships can be considered a precious source of information
about users; by contrast, the approaches mentioned above focus on tagging as the only kind of
information about users. As a further difference, in our approach we analyze tagging behavior at
the {\em semantic level} because we map tags onto concepts through DBPedia/WordNet and we do not limit to
consider only the tagging frequency.

\subsection{Tag Interpretation and Categorization}
\label{sub:tagcategorization}

Recently, some authors introduced the problem of {\em tag categorization}, i.e., they studied how
to classify tags into (often predefined) categories \cite{cantador2011categorising,de2008integrating,de2009exploitation,Schmitz06,WuZhYu06}.

The main motivation behind these work is that many tags are not effective to help users in
retrieving and recommending resources: in fact, sometimes, tags reflect personal opinions/moods of
users (think of tags like ``{\tt beauty}'' or ``{\tt smart}'') and, therefore, they do not provide a relevant
contribution to retrieve contents. Some authors suggested to partition the whole space of tags into
categories; the association of a tag to a category allows to better interpret the meaning of a tag
and decide if it is useful to describe a resource.
An interesting approach to categorizing tags is provided in \cite{cantador2011categorising}.
In that paper, external knowledge bases like WordNet or Wikipedia are used to determine how to map tags onto concepts.
The authors suggest to use heuristics derived from Natural Language Processing to identify tags
expressing subjective facts rather than providing the objective description of a resource.
In \cite{angeletou2009improving} the authors experimentally studied whether WordNet and online ontologies can improve the search activity over a folksonomy.
Experiments showed that WordNet was better than online ontologies to deal with tag synonymies (i.e, different tags sharing the same meaning) and tag polysemy (i.e, the fact that a tag can have different meanings).
By contrast, ontologies were more effective than WordNet in finding objects of interest to the user: in particular, the expansion of a user query with terms coming from an ontology generated a higher number of results than those produced by exploiting terms found in WordNet.
In \cite{pan2009reducing}, the authors describe an approach to supporting search in folksonomies.
In particular, the approach of \cite{pan2009reducing} uses ontologies to deal with tag ambiguity and expand user queries.
Experimental trials reported in \cite{pan2009reducing} are in line with the findings of \cite{angeletou2009improving}, i.e., ontology-based query expansion is actually able of improving the quality of search results:
In addition, ontologies are transparent to the user and this enable non-trained users to use ontologies.

Our work share some similarities with approaches to categorizing tags because we mapped user
contributed tags onto DBPedia concepts to better interpret their meaning. Such a mapping induces a
partitioning of the tags space in which each partition is identified by an ontological concept in
DBPedia.

In \cite{de2008integrating} the authors show how to effectively combine a content-based recommender algorithm with tags. User profiles are learned by applying a multivariate Poisson model exploited for performing text classification. The proposed algorithm works both on static content, i.e., the description of items provided by the Web site and on tags contributed by users to freely annotate
items. The authors observe that tags may suffer from some main syntactic problems like polysemy and synonymy. To deal with these problems the authors suggest to use Natural Language Processing techniques and, in particular, {\em Word Sense Disambiguation} (WSD). The WSD strategy adopted in \cite{de2008integrating} relies on WordNet.

Other approaches rely on Data Mining techniques to find groups of related tags in a folksonomy and
each group identifies a category. For instance, the approach of \cite{WuZhYu06} defines a vectorial
space called {\em conceptual space} whose dimensions represent knowledge categories; each entity of
a folksonomy (i.e., a user, a resource, a tag) is a vector on this conceptual space.
Conceptual space has many applications ranging from the construction of groups of
related tags to the accomplishment of semantic search in a folksonomy.

The approach of \cite{BeKeSm06} first models the tag space as an undirected graph whose nodes
represent tags and whose edges denote co-occurrences of tags in labelling resources. After this graph has been constructed, a \emph{spectral clustering} algorithm is applied to partition it into non-overlapping groups of semantically related tags.

Finally, some authors proposed {\em hierarchical classification schemes} to partition the tag
space. One of the early approaches in this category is presented in \cite{Schmitz06}. In that paper,
the author proposes a probabilistic framework to model subsumption relationships among tags.

Finally, in \cite{de2009exploitation}, the authors describe a technique to build a hierarchy of
tags in a folksonomy. First of all, they define a probabilistic technique to assess if two tags
present related meanings or if the meaning of the first one is more general than that of the second
one. After that, they suggest a greedy algorithm to arrange these tags into a hierarchy.

\section{Conclusions and Future Work}\label{sec:conclusions}
In this article, we investigated tagging and social networking (\emph{friending}) behavior of
individual users across different Social Sharing platforms. We studied the characteristics of user
profiles from Flickr, Delicious and StumbleUpon with respect to three different dimensions: {\em
(i)} intensity of user activities, {\em (ii)} tag-based characteristics of user profiles and {\em
(iii)} semantic characteristics of user profiles. Based on our extensive analysis and the
corresponding findings, we can answer the research questions raised in
Section~\ref{sec:researchquestions} as follows.

\paragraph{Influence of the features of a Social Sharing platform on user behaviors}
We studied to what extent the design and architecture of the
social environment influences the characteristics of the user profiles and to what extent the
personality of a user determine the features of the corresponding profiles.
This led to formulate research questions $Q_1$-$Q_3$ (reported below):

\begin{itemize}

\item $Q_1$. {\em Does the intensity of user activities carried out by a user in a Social
    Sharing system $\mathcal{S}$ depend on the features of the platform?} We studied how much the design features
    of a Social Sharing platform impacted on the social and tagging behaviors of its users. From our analysis, Flickr
    fosters higher involvement in friending activities than Delicious. Moreover, the goal of the tagging functionality has a significant impact on the tagging behavior: for instance, regarding StumbleUpon, the intensity of tagging activities
    is, on average, lower than the intensity of the tagging in Flickr or Delicious. By contrast, this
    difference does not emerge in Flickr and Delicious: in particular, users who are highly
    active in Delicious may also be highly active in Flickr but there is a high fraction of
    users that is highly active in either one of the two environments.

\item $Q_2$. {\em Does the variety of tags exploited by a user depend on the features of the Social Sharing
    system in which she is operating in?} For most users, the variety of the tags they contribute is similar across the different Social Sharing systems. By measuring the entropy of a user's tag-based profiles in the different systems, we
    discovered that the variety of information that is embodied in the profiles depends very
    much on the Social Sharing platform. For the majority of the users, the entropy of the
    Delicious profile is higher than the entropy of StumbleUpon or Flickr profile and this
    makes tagging behavior in Delicious much more unpredictable than that on StumbleUpon or Flickr.

\item $Q_3$. {\em How does the variety of topics a user is concerned with depend on the features and goals of a Social
    Sharing platform?} The semantic analysis of the tag-based profiles proved that the topic
    variety of user profiles clearly differs across Social Sharing platforms. By mapping tags
    onto lexicographic and ontological concepts, we discovered that the variety of the
    lexicographic concepts (i.e. the language that people use to annotate their resources) does
    not strongly vary between environments such as Flickr and Delicious. However, the
    category-like tag suggestions as offered by StumbleUpon result in a lower lexicographic
    variety. When looking at the actual topics that people are concerned with in the different
    Social Sharing platforms (variety of ontological concepts), we see strong differences
    between the environments that complement the tag-based and lexicographic analysis: on
    Flickr we observe a more uniform behavior than on Delicious and StumbleUpon.

\end{itemize}

\paragraph{Correlations between tagging behavior and friending behavior}
We also studied the interplay between tagging and friending activities.
This led to formulate research questions $Q_4$ and $Q_5$ (reported below):

\begin{itemize}

\item $Q_4$. {\em Does the intensity of the activity a user carries out in a Social Sharing
    system depend on the type of activity?} Our analysis showed that there are correlations
    between the intensity of tagging and friending activities. For Flickr, we observe that a
    low tagging activity often implies also a low number of social interactions
(and vice versa) while users who intensively tag show also a
solid intensity regarding their friending activities (and vice versa). By
contrast, in Delicious, the intensity of tagging activities is generally larger than the intensity of friending, which implies
that the architecture of Delicious is not targeted towards social networking activities.

\item $Q_5$. {\em Does the variety of tags a user exploits correlate with the number of
    her social contacts?} On Flickr, the variety of the tags contributed by a user is
    influenced by her friending activities. We identified users who apply a large number of (often
     common) tags (possibly to better promote and increase the visibility of their photos). This behavior may result from the tagging
     design of Flickr, which implies a narrow folksonomy and thus forces users to annotate the
     resources without help of fellow users. On Delicious, which allows for collaborative
     tagging, we did not discover such a correlation.
\end{itemize}

The results of our analysis show how the design and architecture of Social Sharing platforms may
influence the tagging and friending behavior of people. Our results also pose challenges for future
research and pave the way to the development of practical applications.

We plan to extend our cross-system analysis to further systems and types of activities. In
particular, we would like to study whether there exist correlations between microblogging
activities that people perform on Twitter and their tagging and friending activities in Flickr or
Delicious.

A major research avenue we plan to explore is given by the analysis of {\em external factors} and their impact on user behaviors and choices. As external factor we consider, for example, the ``age'' of a system, the system a user joined first and how long a user has been on a system. This would allow us to better understand user behaviors we started analyzing in this paper as well as to formulate new research questions. For instance, could we define and measure the ``loyalty'' of a user? In other words, assume a user joins first Delicious and after some months/weeks Flickr. Our questions are: will she continue spending the largest part of her time on the Web on Delicious or not? Will she spend time equally between Delicious and Flickr? Is there a portion of users who will prefer to dedicate their time to Flickr despite the joined first Delicious? In addition, if a user joins first Delicious and then Flickr, do some differences emerge in her tagging/social behavior? Would such a result change if we would swap Delicious and Flickr (i.e., if we would consider users who join Delicious first and then Flickr)? Is the level of user activities in a platform related to the popularity of the contents hosted in that platform? For instance, if in Flickr are added photos capable of catching the interest of large masses of users, will these new materials increase the level of user participation?

In addition, we are also interested in investigating relationships between static
profile attributes that can be found in Facebook or LinkedIn profiles (such as the profession or
age of a user) and the characteristics of tag-based and social profiles.

Finally, we plan to use tag categorization techniques to improve the results of our analysis.
In particular, tag categorization techniques are useful to identify those tags which support knowledge management and that can be exploited for information retrieval purposes from those reflecting user moods (or their subjective opinions).

\subsubsection*{Acknowledgements}
The research leading to these results has partially received funding from the European Union Seventh
Framework Programme (FP7/2007-2013) in context of the ImREAL
project\footnote{\url{http://imreal-project.eu}}.



\bibliographystyle{acmsmall}
\bibliography{tist2012-bib}

\received{}{}{}

%
%

\end{document}